\documentclass[pra,reprint,a4paper,showpacs]{revtex4-1}
\usepackage{bm}
\usepackage[dvipdfm]{graphicx}
\usepackage{dcolumn}
\usepackage{color}
\usepackage{amsmath}

\newcommand{\ue}[1]{$^{\mathrm{#1}}$}

\draft 

\begin{document}


\title{Time-dependent complete-active-space self-consistent field method
for multielectron dynamics in intense laser fields} 



\author{Takeshi Sato}
\email[Electronic mail:]{sato@atto.t.u-tokyo.ac.jp}

\author{Kenichi L. Ishikawa}
\email[Electronic mail:]{ishiken@atto.t.u-tokyo.ac.jp}

\affiliation{
Photon Science Center, Graduate School of Engineering, The
University of Tokyo, 7-3-1 Hongo, Bunkyo-ku, Tokyo 113-8656, Japan
}



\begin{abstract}
The time-dependent complete-active-space self-consistent-field (TD-CASSCF)
method for the description of multielectron dynamics in intense laser
fields is presented, and a comprehensive description of the method is
given. It introduces the concept of frozen-core (to model tightly bound
electrons with no response to the field), dynamical-core (to model
electrons tightly bound but responding to the field), and active (fully
correlated to describe ionizing electrons) orbital subspaces, allowing
compact yet accurate representation of ionization dynamics in
many-electron systems. The classification into the subspaces can be
done flexibly, according to simulated physical situations and desired
accuracy, and the multiconfiguration time-dependent Hartree-Fock
(MCTDHF) approach is included as a special case. To assess its
performance, we apply the TD-CASSCF method to the ionization dynamics
of one-dimensional lithium hydride (LiH) and LiH dimer models, and
confirm that the present method closely reproduces rigorous MCTDHF
results if active orbital space is chosen large enough to include
appreciably ionizing electrons. The TD-CASSCF method will open a way to
the first-principle theoretical study of intense-field induced
ultrafast phenomena in realistic atoms and molecules. 
\end{abstract}


\maketitle 

\section{introduction\label{sec:introduction}}
The advent of the chirped pulse amplification (CPA) technique
\cite{Strickland1985OC} has enabled the production of femtosecond laser
pulses whose focused intensity easily exceeds $10^{14}\,{\rm W/cm}^2$
and even reach $\sim 10^{22}\,{\rm W/cm}^2$ 
\cite{Bahk2004OL,Yanovsky2008OE,Yu2012OE}. Exposed to
visible-to-mid-infrared pulses with an intensity  
typically higher than $10^{14}\,{\rm W/cm}^2$,
atoms and molecules exhibit nonperturbative nonlinear response such as
above-threshold ionization (ATI), tunneling ionization,  
high-order harmonic generation (HHG), and nonsequential double
ionization (NSDI) \cite{Protopapas1997RPP,Brabec2000RMP}. 
HHG, for example, represents a highly successful avenue
toward an ultrashort coherent light source in the  
extreme-ultraviolet (XUV) and soft x-ray regions
\cite{Seres2005Nature,Chang}. The development of these novel light
sources has opened new research possibilities including ultrafast
molecular probing
\cite{Itatani2004Nature,Haessler2010NPhys,Salieres2012RPP}, attosecond
science \cite{Agostini2004RPP,Krausz2009RMP,Gallmann2013ARPC}, and XUV
nonlinear optics \cite{Sekikawa2004Nature,Nabekawa2005PRL}. 

In parallel with the progress in experimental techniques, various
numerical methods have been developed to explore atomic and molecular
dynamics in intense laser fields. While direct solution of the time-dependent
Schr\"odinger equation (TDSE) provides exact description, this method is
virtually unfeasible for multielectron systems beyond He
\cite{Pindzola1998PRA,Pindzola1998JPB,Colgan2001JPB, 
Parker2001,Laulan2003PRA,Piraux2003EPJD,Laulan2004PRA,ATDI2005,
Feist2009PRL,Pazourek2011PRA,He_TPI2012PRL,Suren2012PRA,He_TPI2013AS}
and ${\rm H}_2$ \cite{Vanroose2006PRA,Horner2008PRL,Lee2010JPB}. As a
result, single-active electron (SAE) approximation is widely used, in
which only the outermost electron is explicitly treated, and the effect
of the others, assumed to be frozen, is embedded in a
model potential. This approximation, however, fails to account for
multielectron and multichannel effects
\cite{Haessler2010NPhys,Gordon2006PRL,Rohringer2009PRA,Smirnova2009PNAS,Akagi2009Science,Boguslavskiy2012Science}
in high-field phenomena. 
Thus, alternative many-electron methods are
required to catch up with new experimental possibilities. For example,
Caillat {\it et al.} \cite{Caillat2005PRA} have introduced the
multiconfiguration time-dependent Hartree-Fock (MCTDHF) approach (see
below) to study correlated multielectron systems in strong laser
fields. Although they have presented the results for 
up to six-electron model molecules, 
its computational time {\rm prohibitively}
increases with the number of electrons. 
Another interesting route is the
time-dependent configuration-interaction singles (TD-CIS) method,
implemented by Santra and coworkers.
\cite{Rohringer2006PRA,Greenman2010PRA}, in which the many-electron
wavefunction is expanded in terms of the Hartree-Fock (HF) 
ground state and singly excited configuration state functions (CSF).
The TD-CIS method has an advantage to give a clear one-electron
picture for multichannel ionizations. However,
its applications are limited to the dynamics dominated by single
ionizations, with an initial state described correctly by the HF method.
%
Time-dependent density functional theory (TDDFT)
\cite{Gross1996TCC,Otobe2004PRA,Telnov2009PRA}, though attractive for
its low computational cost, delivers only the electron density, not the
wavefunction, rendering the definition of observables difficult. More
seriously, it is difficult to estimate and systematically improve the
accuracy of the exchange-correlation potential. 

Among the more recent developments, 
the orbital-adapted time-dependent coupled-cluster (OATDCC) method
proposed by Kvaal \cite{Kvaal:2012} is of particular interest, which
pioneered the time-dependent coupled-cluster (CC) approach with
biviriationally adapted orbital functions and excitation amplitudes. 
The fixed-orbital CC method was also implemented by Huber and Klamroth
\cite{Huber:2011}. 
The time-dependent CC approach should be a promising avenue to the
time-dependent many-electron problems in view 
of the spectacular success of the stationary CC theory. However, it
seems to require further theoretical sophistications to make a rigorous
numerical method on it.
Hochstuhl and Bonitz proposed the time-dependent restricted-active-space
configuration-interaction (RASCI) method \cite{Hochstuhl:2012}, in which
the total wavefunction is expanded with fixed Slater determinants
compatible to the RAS constraints known in quantum chemistry
\cite{Olsen:1988}. 
The TD-RASCI method, with cleverly devised spatial partitioning, has been
successfully applied to helium and beryllium atoms \cite{Hochstuhl:2012}.
A disadvantage of the method is the lack of the size extensivity
\cite{Szabo:1996, Helgaker:2002} 
discussed in Sec.~\ref{sec:results}, which is a common problem of
truncated CI approaches, except the simplest TD-CIS
\cite{Rohringer2006PRA}.

In this work, we propose a flexible {\it ab initio} time-dependent
many-electron method based on the concept of the complete
active-space self-consistent field (CASSCF) 
\cite{Ruedenberg:1982, Roos:1980, Roos:1987, Schmidt:1998}.
Our approach is derived
from the first principles, and simultaneously, fills the huge gap between
the MCTDHF method and SAE approximation.

Most of the ground-state closed-shell systems are described
qualitatively well by the HF method \cite{Szabo:1996}. However, the time-dependent
Hartree-Fock (TDHF) method cannot describe ionization processes \cite{Pindzola:1991}, since
it enforces to keep the initial closed-shell structure. Instead, at
least two orbital functions are required to describe the field
ionization of a singlet two-electron system. The spatial
part of the total wavefunction in such approximation reads
\begin{subequations}
\label{eq:gvb}
\begin{eqnarray}
\label{eq:gvb_ehf}
\Psi(1,2) &\propto& \psi_1(1)\psi_2(2) + \psi_2(1)\psi_1(2) \\ &=&
\label{eq:gvb_no}
C_1\phi_1(1)\phi_1(2) + C_2\phi_2(1)\phi_2(2),
\end{eqnarray}
\end{subequations}
The first form is known as the extended Hartree-Fock (EHF)
\cite{Pindzola:1997, Tolley:1999, Dahlen:2001}
wavefunction, and has been successfully applied to the
intense-field phenomena for two-electron systems \cite{Pindzola:1997,
Dahlen:2001, Nguyen:2006}.
The second form is obtained by the canonical orthogonalization of
non-orthogonal orbitals $\psi_{i=1,2}$ 
\cite{Szabo:1996}, 
and an example of the configuration-interaction (CI) wavefunction 
\cite{Szabo:1996, Helgaker:2002}. 
It is clear that not only 
the CI coefficients $\left\{C_i(t)\right\}$ but also the orbitals $\left\{\phi_i(t)\right\}$
have to be varied in time in order to properly describe the ionization. Thus we
need the multiconfiguration Hartree-Fock (MCHF) or the
multiconfiguration self-consistent field (MCSCF) wavefunction 
\cite{Szabo:1996, Schmidt:1998, Helgaker:2002}, 
where both the CI coefficients and the shape of the orbitals are the
variational degrees of freedom.

This idea has been realized for many-electron systems by the
MCTDHF method \cite{Caillat2005PRA, Kato:2004}, 
in which the total wavefunction is expanded in terms of
Slater determinant (or CSF) bases,
\begin{eqnarray}
\label{eq:mcscf}
 \Psi(t) = \sum_I C_I(t) \Phi_I(t),
\end{eqnarray}
where both CI coefficients $\left\{C_I(t)\right\}$
and bases $\left\{\Phi_I(t)\right\}$ are allowed to vary in time.
The Slater determinants are constructed
(in the spin-restricted treatment, see Sec.~\ref{subsec:mcscf}) 
from $2n$ spin-orbitals $\left\{\phi_p; p=1,2,\cdot\cdot
n\right\} \otimes \left\{\alpha,\beta\right\}$ , where
$\left\{\phi_p(t)\right\}$ are time-dependent spatial orbital functions
and $\alpha$ ($\beta$) is the up- (down-) spin eigenfunction.
See also 
Refs.~\cite{Nest:2005a, Nest:2005b, Jordan:2006, Kato:2008,
Alon:2007, Hochstuhl:2011, Miranda:2011a}
for the MCTDHF method, and Beck {\it et al} \cite{Beck:2000} and
references therein for the precedent multiconfiguration time-dependent
Hartree method for Bosons.

Despite its naming,
which in principle refers to the {\it general} multiconfiguration
wavefunction of the form Eq.~(\ref{eq:mcscf}) with flexible
choice of expansion bases 
(range of summation $I$), 
previous implementations of the MCTDHF method (except the fixed-CI
formulation of Ref.~\cite{Miranda:2011a}) were limited to the full-CI
expansion; 
the summation $I$ 
of Eq.~(\ref{eq:mcscf}) 
is over all the possible ways of
distributing $N$ electrons among the $2n$ spin-orbitals.
More intuitively, we write such an MCTDHF wavefunction symbolically as
\begin{eqnarray}
\label{eq:fullci_symbolic} 
\Psi_{\rm MCTDHF} : \left\{
\phi_1(t)\phi_2(t)...\phi_n(t)
\right\}^N,
\end{eqnarray}
which is understood to represent the $N$-electron full-CI wavefunction
using $n$ time-dependent orbitals.
Though powerful, the MCTDHF method suffers from severe limitation in
the applicability to large systems, since the full-CI dimension grows
factorially with increasing $N$. 

It is reasonable to expect that in a large molecule interacting with
high-intensity, long-wavelength lasers, the deeply bound electrons
remain non-ionized, while only the higher-lying valence electrons ionize
appreciably. For the bound electrons, a closed-shell description of the
HF type would be acceptable as a first approximation \cite{Li:2005}.
On the other hand, correlated treatment is required
for ionized electrons to describe the seamless transition from the
closed-shell-dominant initial state into the symmetry-breaking continuum
(discussed in Sec.~\ref{subsec:1d-lih}).

\begin{figure}[!b]
\includegraphics[width=0.4\textwidth,clip]{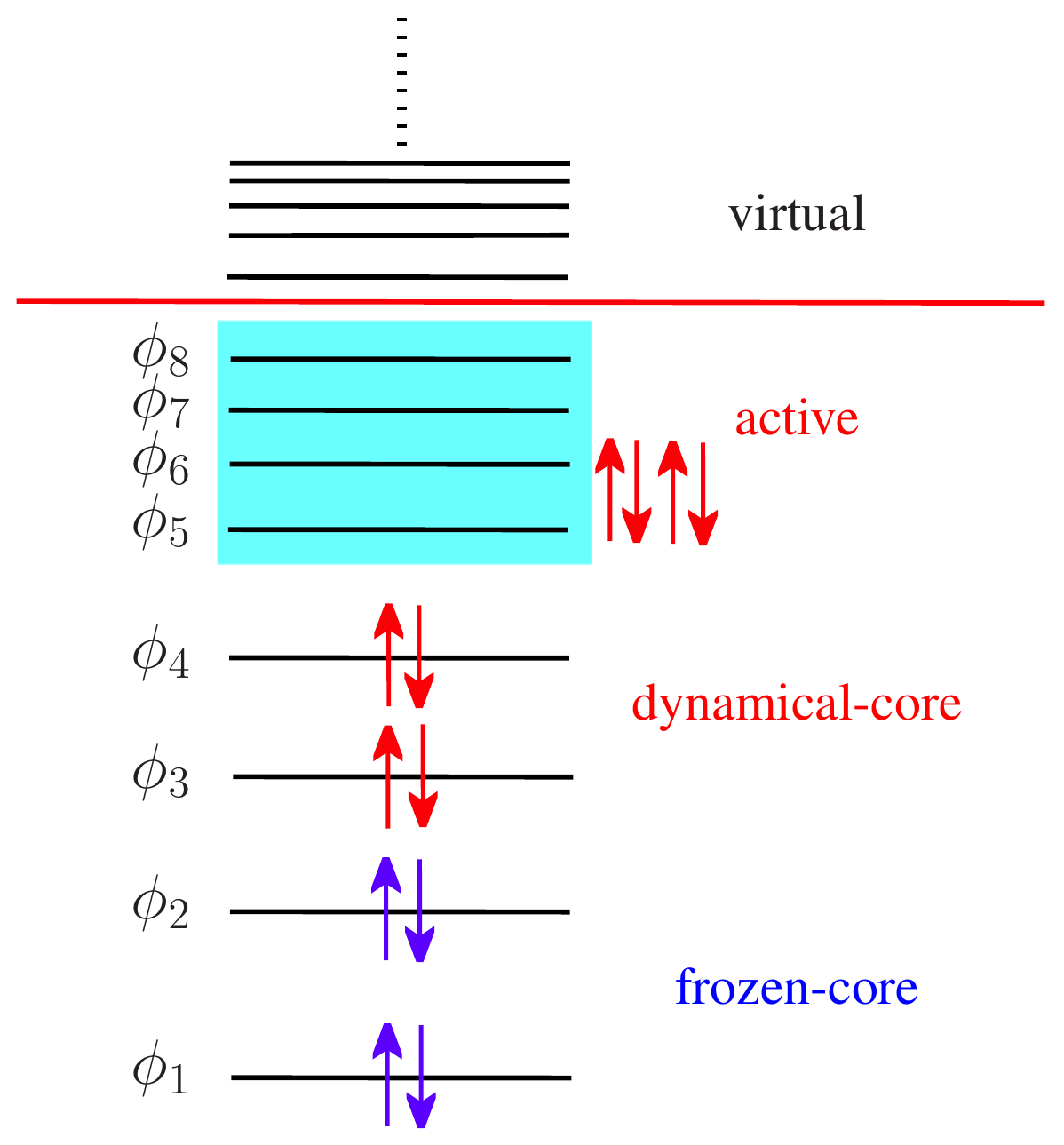}
\caption{\label{fig:td-casscf}Pictorial explanation of the TD-CASSCF
 concept, illustrating a 12-electron system with two frozen-core
 (orbitals 1 and 2), two dynamical-core (3 and 4), and four active
 orbitals (5 to 8). Solving the orbital-EOM guarantees the variational
 splitting, in a time-dependent sense, of core-active, core-virtual, and
 active-virtual orbital subspaces.}%
\end{figure}

The CASSCF method 
\cite{Ruedenberg:1982, Roos:1980, Roos:1987, Schmidt:1998}
provides an ideal {\it ansatz} for such a
problem. It introduces the concept of {\it core} and {\it active}
orbital subspaces, and spatial orbitals participating in Eq.~(\ref{eq:fullci_symbolic})
are classified into these subspaces. The core orbitals are forced to be
doubly occupied all the time, while the active orbitals are allowed
general (0, 1, or 2) occupancies. Thus, the CASSCF wavefunction is written
symbolically as
\begin{subequations}
\label{eq:casci_symbolic} 
\begin{eqnarray}
\Psi_{\rm CASSCF}(t) &:& \phi^2_1 \phi^2_2 \cdot\cdot\cdot
\phi^2_{n_{\rm C}} \label{eq:casci_symbolic_core} \\ &\times&
\left\{
\phi_{n_{\rm C} + 1} \phi_{n_{\rm C} + 2} \cdot\cdot\cdot \phi_{n_{\rm C} + n_{\rm A}}
\right\}^{N_{\rm A}}, \label{eq:casci_symbolic_active} 
\end{eqnarray}
\end{subequations}
where factors (\ref{eq:casci_symbolic_core}) and
(\ref{eq:casci_symbolic_active}) represent core- and active-subspaces,
respectively, and the total wavefunction is properly antisymmetrized. 
See Sec.~\ref{subsec:mcscf} for the rigorous definition.
The $n_{\rm C}$ {\it core orbitals} describe $N_{\rm C} = n_{\rm C} / 2$
{\it core electrons} within the closed-shell constraint, while the 
$N_{\rm A}$ {\it active electrons} are fully correlated using $n_{\rm A}$
{\it active orbitals}.
Whereas, in general, all the orbitals are varied in time, it is also possible to
further split the core space into {\it frozen-core} (fixed) and
{\it dynamical-core} (allowed to vary in time in response to the field)
subspaces. See Fig.~\ref{fig:td-casscf}, which illustrates the concept
of the orbital subspacing.
%

The equation of motion (EOM) of our method, called TD-CASSCF,
is derived 
based on the time-dependent variational principle (TDVP)
\cite{Frenkel:1934, Lowdin:1972, Moccia:1973}, 
as detailed in Sec.~\ref{sec:theory}. 
It guarantees the best possible solution using the total wavefunction
expressed as Eq.~(\ref{eq:casci_symbolic}). 
The fully correlated active space allows to describe ionization
processes which involve the strong correlation due to the breaking
closed-shell symmetry (discussed in Sec.~\ref{subsec:1d-lih}).
It can also include multichannel and multielectron effects
(discussed in Sec.~\ref{subsec:1d-lihd}).
The dynamical core orbitals account for the effect of field-induced
core polarizations. In whole, the TD-CASSCF method enables
compact yet accurate representation of multielectron dynamics, if the
active space is chosen correctly according to the physical processes of
interest.


This paper proceeds as follows. 
In Sec.~\ref{sec:theory}, the details of the TD-CASSCF method are described.
Then in Sec.~\ref{sec:results}, the TD-CASSCF method is numerically
assessed for ionization dynamics of one-dimensional multielectron
models. Finally, Sec.~\ref{sec:conclusions} concludes this work and
discusses future prospects. Appendix~\ref{app:ionp} provides the
definition and computational details for real-space domain-based
ionization probabilities.
The Hartree atomic units are used throughout unless otherwise noted.

\section{Theory
\label{sec:theory}}
In this section, we derive the EOMs for the TD-CASSCF method.
To this end, first we give the rigorous definitions of the MCTDHF, MCSCF,
CASSCF, and general multiconfiguration
{\it ansatz} for the total 
wavefunction in Sec.~\ref{subsec:mcscf}.
Next in Sec.~\ref{subsec:g-tdmcscf}, the EOMs
of orbitals and CI coefficients for the general
multiconfiguration wavefunctions 
are discussed by reviewing the work of Miranda {\it
et al} \cite{Miranda:2011a}.
Then we specialize the general formulation to the TD-CASSCF method in
Sec.~\ref{subsec:td-casscf} to derive the explicit EOMs, and discuss its
computational aspects in Sec.~\ref{subsec:theory_remarks}. 

\subsection{Multiconfiguration wavefunctions
\label{subsec:mcscf}}
Our formulation is determinant-based 
\cite{Olsen:1988, Hochstuhl:2011}
within the spin-restricted
treatment, i.e., using the same spatial orbitals for up- and down-spin electrons.
We define $n$ {\it occupied} spatial orbitals
$\{\phi_p; p=1,2,\cdot\cdot\cdot,n\}$ and 
$N_b - n$ {\it virtual} orbitals 
$\{\phi_a; a=n+1,n+2,\cdot\cdot\cdot,N_b\}$, here $N_b$ is the
dimension of the spinless one-particle Hilbert space, determined by
e.g., the number of grid points to discretize the orbitals or the number
of intrinsic single-particle basis functions to expand the orbitals.
The indices $\{p,q,r,s\}$, $\{a,b\}$, and
$\{\mu,\nu,\lambda,\gamma,\delta\}$ are used to label occupied, virtual,
and general (occupied + virtual) orbitals, respectively, and $\{\sigma,
\tau\} \in \{\alpha, \beta\}$ label spin eigenfunctions.
In the followings, Einsteins's summation convention is applied for
repeated upper and lower orbital indices within a term, with summation
ranges implicit as above. 
The orbitals are assumed to be orthonormal all the time, 
\begin{eqnarray}\label{eq:orthonormality}
\langle \phi_\mu(t) \vert \phi_\nu(t) \rangle \equiv 
\int\!d\bm{r} \phi^*_\mu(t) \phi_\nu(t) = \delta^\mu_\nu,
\end{eqnarray}
where $\delta^\mu_\nu$ denotes the Kronecker delta. The Slater
determinants $\Phi_{\rm I} (t)$ in Eq.~(\ref{eq:mcscf}) are constructed, as usual,
from the occupied orbitals $\{\phi_p\}$, and given in the
occupation number representation 
\cite{Helgaker:2002, Hochstuhl:2011} as
\begin{eqnarray}
\label{eq:onv}
 \vert I(t) \rangle &=&
 \vert I^\alpha_1 I^\alpha_2 \cdot\cdot\cdot I^\alpha_n
       I^\beta_1 I^\beta_2 \cdot\cdot\cdot I^\beta_n; t \rangle
 \nonumber \\
 &=& \prod_{p = 1}^n 
 (\hat{a}^\dagger_{p\alpha})^{I^\alpha_p}
 (\hat{a}^\dagger_{p\beta})^{I^\beta_p} \vert vac \rangle,
\end{eqnarray}
where $\hat{a}^\dagger_{\mu\sigma}$ (and $\hat{a}_{\mu\sigma}$ appearing
below) is the Fermion creation (destruction) operator for 2$N_b$
spin-orbitals $\{\phi_\mu\} \otimes \{\alpha,\beta\}$.
The integer array $\{I^\sigma_p = 0, 1\}$ specifies the occupancy of each
spin-orbital, where $\sum_{p=1}^n I^\sigma_p = N^\sigma$, with
$N^\sigma$ being the number of $\sigma$-spin electrons, and $N =
N^\alpha + N^\beta$.
The time-dependence of the occupation number vector $\vert I(t)
\rangle$ is implicit in the spatial orbitals.

We focus on the dynamics induced by spin-independent external fields,
and the initial wavefunction is assumed to be the spin
eigenfunction. Therefore the total and projected spin operators,
$\hat{S}^2, \hat{S}_z$ are the constants of motion. Each Slater
determinant is the eigenfunction of $\hat{S}_z$ with eigenvalue
$N^\alpha - N^\beta$, while not generally of $\hat{S}^2$. The total
wavefunction is spin-adapted, however, since the initial state is prepared by the
variational optimization of CI coefficients, which automatically gives
proper spin combinations.

Throughout this paper, the term MCTDHF is used for the method based on
the full-CI expansion using $n$ occupied orbitals:
\begin{eqnarray}
\label{eq:mctdhf_2q}
 \vert \Psi_{\rm MCTDHF}(t) \rangle = 
 \sum_I^{\Pi_{\rm FCI}} C_I(t) \vert I(t) \rangle,
\end{eqnarray}
with $I$ varying freely in the full-CI space $\Pi_{\rm FCI}$,
spanned by all the determinants generated from the 2$n$ occupied spin-orbitals.
This is the current standard of the MCTDHF method as mentioned in
Sec.~\ref{sec:introduction}. Note that Miranda {\it et al}
\cite{Miranda:2011a} used the term MCTDHF in a broader sense to denote
approaches based on the {\it general} multiconfiguration wavefunctions. 
To be definite, we call the general {\it ansatz} as MCSCF:
\begin{eqnarray}
\label{eq:mcscf_2q}
 \vert \Psi_{\rm MCSCF}(t) \rangle = 
 \sum_I^{\Pi} C_I(t) \vert I(t) \rangle,
\end{eqnarray}
with the general CI space $\Pi$ defined as any arbitrary subspace of $\Pi_{\rm
FCI}$, $\Pi = \{ \vert I \rangle \in \Pi_{\rm FCI}; C_I(t)
\equiv\hspace{-1.1em}/ \hspace{.5em} 0 \}$. 
A trivial example of this class is the single-determinant HF
wavefunction for closed-shell singlet or
open-shell high-spin states.
The only nontrivial applications of Eq.~(\ref{eq:mcscf_2q}) to the
time-dependent problems made thus far is the general open-shell TDHF
approaches formulated in Ref.~\cite{Miranda:2011a}, in which the CI
coefficients are determined by the spin-symmetry and time-independent.

The most successful MCSCF method in quantum chemistry is the
CASSCF (also known as fully optimized reaction space) method
introduced in Sec.~\ref{sec:introduction} [Eq.\ (\ref{eq:casci_symbolic})], in which
the CI expansion is limited to the space spanned by Slater determinants
that include $n_{\rm C}$ doubly occupied core orbitals, called the CASCI space
$\Pi_{\rm CAS}$:
\begin{eqnarray}
\label{eq:casscf_2q}
 \vert \Psi_{\rm CASSCF}(t) \rangle = \sum_I^{\Pi_{\rm CAS}} C_I(t)
 \vert I(t) \rangle,
\end{eqnarray}
\begin{eqnarray}
\label{eq:determinant_casci}
 \vert I \rangle = 
 \prod_{i = 1}^{n_{\rm C}}
 \hat{a}^\dagger_{i\alpha}
 \hat{a}^\dagger_{i\beta}
 \prod_{t = n_{\rm C} + 1}^n
 (\hat{a}^\dagger_{t\alpha})^{I^\alpha_t}
 (\hat{a}^\dagger_{t\beta})^{I^\beta_t} \vert vac \rangle,
\end{eqnarray}
with $\sum_t I^\sigma_t = N^\sigma_{\rm A}$ being the number of
$\sigma$-spin active electrons satisfying $N^\sigma_{\rm A} = N^\sigma -
N_{\rm C}/2$ and $N_{\rm A} = N^\alpha_{\rm A} + N^\beta_{\rm A}$.
Hereafter, we use orbital indices
$\{i,j,k,l\}$ for core- and $\{t,u,v,w,x\}$ for active-orbitals, while
keeping $\{p,q,r,s\}$ for general occupied (core + active) orbitals.
Following the convention in the electronic structure theory
\cite{Szabo:1996, Helgaker:2002}, 
we use the acronym CASSCF($N_{\rm A}, n_{\rm A}$)
to denote the CASSCF wavefunction with $N_{\rm A}$ active electrons
and $n_{\rm A}$ active orbitals. The MCTDHF wavefunction with $n$ occupied
orbitals is identical to CASSCF($N, n$) and denoted as MCTDHF($n$).
See Fig.~\ref{fig:td-casscf},
Eqs.~(\ref{eq:fullci_symbolic}) and (\ref{eq:casci_symbolic}) in
Sec.~\ref{sec:introduction}, and
Eqs.~(\ref{eq:cas_for_lih}) and (\ref{eq:cas_for_lihd}) in
Sec.~\ref{sec:results} for intuitive understanding of these notations.

\subsection{Equations of motion for MCSCF wavefunctions
\label{subsec:g-tdmcscf}}
Recently, Miranda {\it et al} discussed EOMs for MCSCF
wavefunctions \cite{Miranda:2011a}.
Although their main motivation was the fixed-CI formulations,
they also presented important equations applicable to the general MCSCF
wavefunctions (See Sec.~IV of Ref.~\cite{Miranda:2011a}).
Here we follow the essentials of their development to obtain
Eqs.~(\ref{eq:g-tdci}) and (\ref{eq:g-tdmo}) below.

The spin-free second-quantized Hamiltonian is given by
\begin{eqnarray}
\label{eq:hamiltonian}
\hat{H} = h^\mu_\nu \hat{E}^\mu_\nu
+ \frac{1}{2} g^{\mu\lambda}_{\nu\gamma} \hat{E}^{\mu\lambda}_{\nu\gamma},
\end{eqnarray}
where $h^\mu_\nu$ and $g^{\mu\lambda}_{\nu\gamma}$ are the one- and two-electron Hamiltonian
matrix elements,
\begin{eqnarray}
\label{eq:ham1e}
h^\mu_\nu = \int d\bm{r} \phi^*_\mu(\bm{r}) 
h\left(\bm{r}, \bm{\nabla}_r\right)
\phi_\nu(\bm{r}),
\end{eqnarray}
\begin{eqnarray}
\label{eq:ham2e}
g^{\mu\lambda}_{\nu\gamma} = \int\!\!\!\int \! d\bm{r}_1 d\bm{r}_2 &&
\phi^*_\mu(\bm{r}_1) \phi^*_\lambda(\bm{r}_2)
V_{ee}(\bm{r}_1, \bm{r}_2)  \nonumber \\ \times &&
\phi_\nu(\bm{r}_1) \phi_\gamma(\bm{r}_2), 
\end{eqnarray}
with $h$ consisting of kinetic, nucleus-electron, and external laser
terms, and $V_{ee}$ being the electron-electron interaction, and
\begin{eqnarray}
\label{eq:e1}
\hat{E}^\mu_\nu = \sum_\sigma
\hat{a}^\dagger_{\mu\sigma}\hat{a}_{\nu\sigma},
\end{eqnarray}
\begin{eqnarray}
\label{eq:e2}
\hat{E}^{\mu\lambda}_{\nu\gamma} = \sum_{\sigma\tau}
\hat{a}^\dagger_{\mu\sigma}\hat{a}^\dagger_{\lambda\tau}
\hat{a}_{\gamma\tau}\hat{a}_{\nu\sigma} =
\hat{E}^\mu_\nu\hat{E}^\lambda_\gamma - \hat{E}^\mu_\gamma \delta^\lambda_\nu.
\end{eqnarray}

Following the TDVP
\cite{Frenkel:1934, Lowdin:1972, Moccia:1973},
the action integral $S$,
\begin{eqnarray}
\label{eq:action}
 S[\Psi] = \int_{t_0}^{t_1} \! dt 
\langle \Psi \vert 
\left(\hat{H} - {\rm i}\frac{\partial}{\partial t}\right)
\vert \Psi \rangle,
\end{eqnarray}
is made stationary,
\begin{eqnarray}
\label{eq:delta_action}
\delta S = \int_{t_0}^{t_1} \! dt &&
\left\{
\langle \delta\Psi \vert 
\left(\hat{H} \vert \Psi \rangle - {\rm i} \vert \dot{\Psi} \rangle \right)
\right. \nonumber \\ + && \left. 
\left( \langle \Psi \vert \hat{H} + {\rm i}\langle \dot{\Psi} \vert \right)
\vert \delta\Psi \rangle
\right\} = 0, 
\end{eqnarray}
with respect to allowed variations $\delta \Psi$ of the total wavefunction,
where $\dot{\Psi} \equiv \partial \Psi / \partial t$. 
The time derivative of
$\delta\Psi$ is integrated out by part, assuming $\delta\Psi(t_0) =
\delta\Psi(t_1) = 0$. See Ref.~\cite{Moccia:1973} for the formal discussion
on the validity of this procedure.
By taking the orbital orthonormality
into account, the variations and the time derivatives of an orbital
$\phi_\mu$ can be written as
$\delta\phi_\mu = \phi_\nu \Delta^\nu_\mu$, and
${\rm i}\dot{\phi}_\mu = \phi_\nu R^\nu_\mu$, respectively, 
with $R^\mu_\nu \equiv {\rm i}\langle \phi_\mu \vert \dot{\phi}_\nu \rangle$.
The matrix $\Delta$ is anti-Hermitian, while $R$ is Hermitian \cite{Miranda:2011a}.
Then, the allowed variation and the time derivative of the total
wavefunction are compactly given by
\begin{eqnarray}
\label{eq:delta_mcscf}
  \vert\delta\Psi\rangle = \sum_I^{\rm \Pi} \vert I\rangle
  \delta C_I +
  \hat{\Delta} \vert\Psi\rangle,
\end{eqnarray}
\begin{eqnarray}
\label{eq:dt_mcscf}
  {\rm i}\vert\dot{\Psi}\rangle = 
  {\rm i}\sum_I^{\rm \Pi} \vert I\rangle \dot{C}_I +
  \hat{R} \vert\Psi\rangle, 
\end{eqnarray}
where $\delta C_I$ and $\dot{C}_I$ are the variation and the time derivative
of CI coefficient $C_I$, and $\hat{\Delta} = \Delta^\mu_\nu \hat{E}^\mu_\nu$,
$\hat{R} = R^\mu_\nu \hat{E}^\mu_\nu$.
Inserting Eqs.~(\ref{eq:dt_mcscf}) and (\ref{eq:dt_mcscf}) and their
Hermitian conjugates $\langle \delta\Phi \vert$ and $-{\rm i}\langle
\dot{\Phi}\vert$ into Eq.~(\ref{eq:delta_action}), and
requiring the equality for individual variations $\delta C_I(t)$ and
$\Delta^p_q(t)$, after some algebraic manipulations \cite{Miranda:2011a} we have
\begin{eqnarray}
\label{eq:g-tdci}
 {\rm i}\dot{C}_I = \langle I\vert \bar{H} \vert \Psi\rangle,
\end{eqnarray}
\begin{eqnarray}
\label{eq:g-tdmo}
\langle \Psi \vert \bar{H} (1 - \hat{\Pi}) \hat{E}^\mu_\nu \vert \Psi 
\rangle -
\langle \Psi \vert \hat{E}^\mu_\nu (1 - \hat{\Pi}) \bar{H} \vert \Psi
\rangle = 0,
\end{eqnarray}
where $\bar{H} = \hat{H} - \hat{R}$, and $\hat{\Pi} = \sum_{I \in \Pi}
\vert I\rangle\langle I\vert$ is the configuration projector onto the
general CI space $\Pi$. The system of equations, Eqs.~(\ref{eq:g-tdci})
and (\ref{eq:g-tdmo}), is to be solved for ${\rm i} \dot{C}_I$ and $R$,
which determine the time-dependence of CI coefficients and orbitals,
respectively.
In Ref.~\cite{Miranda:2011a}, these equations appeared as an intermediate to
derive the MCTDHF equation, rather than as the final result.
Here we emphasize that Eqs.~(\ref{eq:g-tdci}) and (\ref{eq:g-tdmo})
are valid for {\it general} MCSCF wavefunction fit into the form 
of Eq.~(\ref{eq:mcscf_2q}).
Equation~(\ref{eq:g-tdmo}) is also extensively discussed by Miyagi and
Madsen in their recent development of MCTDHF method with restricted CI
expansions \cite{Miyagi:2013}.

\subsection{TD-CASSCF equations of motion
\label{subsec:td-casscf}}
\subsubsection{Orbital equations of motion
\label{subsec:td-casscf-mo}}
Now we apply the CASSCF constraint defined in Sec.~\ref{subsec:mcscf}
to the general orbital-EOM derived in Sec.~\ref{subsec:g-tdmcscf}.
Equation~(\ref{eq:g-tdmo}), with $\hat{\Pi}$ replaced by $\hat{\Pi}_{\rm
CAS}$, reduces to a trivial identity for an orbital pair
$\{\mu,\nu\}$ belonging to a same orbital subspace
(core, active, or virtual), 
since the singly replaced determinants,
$\hat{E}^i_j\vert I\rangle = 2\delta^i_j \vert I\rangle$, 
$\hat{E}^t_u\vert I\rangle$, or
$\hat{E}^a_b\vert I\rangle = 0$,
either fall within $\Pi_{\rm CAS}$ or vanish,
and the configuration projector $1-\hat{\Pi}_{\rm CAS}$ eliminates such
contributions. We refer to these intra-subspace orbital rotations 
$\{\hat{E}^i_j, \hat{E}^t_u, \hat{E}^a_b\}$ as {\it redundant}, since
the total wavefunction is invariant under such orbital
transformations, if accompanied by the corresponding transformation 
of CI coefficients \cite{Helgaker:2002, Caillat2005PRA}.

The redundant orbital rotations can be excluded in varying our action
functional in Eq.~(\ref{eq:delta_mcscf}), since their effects 
to $\delta \Psi$ are taken into account by the CI variations $\delta
C_I$.
On the other hand, for $\{\mu,\nu\}$ belonging to 
different orbital subspaces (core-active, core-virtual, or active-virtual),
the projector $\hat{\Pi}_{\rm CAS}$ can be dropped in
Eq.~(\ref{eq:g-tdmo}), and we have a simpler expression, 
\begin{eqnarray}
\label{eq:qg-tdmo}
\langle \Psi \vert \left[\hat{H} - \hat{R}, \hat{E}^\mu_\nu\right] \vert
\Psi \rangle = 0,
\end{eqnarray}
with $\hat{E}^\mu_\nu = \{\hat{E}^a_p, \hat{E}^p_a, \hat{E}^t_i,
\hat{E}^i_t\}$, constituting the {\it non-redundant} orbital rotations.
The general orbital-EOM of Eq.~(\ref{eq:g-tdmo}) is thus reduced to
Eq.~(\ref{eq:qg-tdmo}), which is to be solved only for the non-redundant
orbital pairs.

It is fascinating to see an analogy in Equation~(\ref{eq:qg-tdmo}) with
the time-independent MCSCF theory; It is formally 
identical to the generalized Brillouin condition of the stationary
wavefunction \cite{Levy:1968, Levy:1969}, if we replace $\hat{H} - \hat{R}$ with $\hat{H}$.
Thus, the remaining derivations are parallel to the time-independent theory.
We can explicitly write down the matrix elements of Eq.~(\ref{eq:qg-tdmo}) to obtain
\begin{eqnarray}\label{eq:qg-tdmcscf}
  R^\mu_\lambda D^\lambda_\nu -
  D^\mu_\lambda R^\lambda_\nu 
  =
  F^\mu_\nu - F^{\nu*}_\mu,
\end{eqnarray}
\begin{eqnarray}\label{eq:qg-fock}
 F^\mu_\nu = h^\mu_\lambda D^\lambda_\nu + 
 g^{\mu\gamma}_{\lambda\delta}
 P^{\lambda \delta}_{\nu \gamma}, 
\end{eqnarray}
where $D^\mu_\nu \equiv \langle\Psi\vert\hat{E}^\nu_\mu\vert\Psi\rangle$ and 
$P^{\mu\lambda}_{\nu\gamma} \equiv
\langle\Psi\vert\hat{E}^{\nu\gamma}_{\mu\lambda}\vert\Psi\rangle$ are
one- and two-electron reduced density matrix (RDM) elements, respectively.
The matrix $F$ is called the generalized Fock matrix, whose Hermiticity, leading to
vanishing right-hand side of Eq.~(\ref{eq:qg-tdmcscf}), is the
stationary condition with respect to the orbital variations.
\cite{Roos:1980, Roos:1987, Schmidt:1998, Helgaker:2002}

The nonzero density matrix elements of the CASSCF wavefunction are
$D^i_j = 2\delta^i_j, D^t_u, P^{ij}_{kl} = 4\delta^i_k\delta^j_l -
2\delta^i_l \delta^j_k,
P^{ti}_{ui} = 2D^t_u, P^{ti}_{iu} = -D^t_u$ and $P^{tu}_{vw}$.
Then the required generalized Fock matrix elements read \cite{Roos:1980}
\begin{eqnarray}
\label{eq:qg-fock-core}
 F^\mu_i = 2(f^\mu_i + G^\mu_i),
\end{eqnarray}
\begin{eqnarray}
\label{eq:qg-fock-active}
 F^\mu_t = f^\mu_u D^u_t 
  + \left({\mit \Gamma}_t\right)^\mu_t,
\end{eqnarray}
where the matrices $f$, $G$, and ${\mit \Gamma}_t$ represent, respectively,
operators $\hat{f}$, $\hat{G}$, and $\hat{\mit \Gamma}_t$ given by
\begin{eqnarray}
\label{eq:core-fock}
 \hat{f} = \hat{h}^{\rm FC} + \sum_j^{\rm d.c.} \left(2\hat{J}_j - \hat{K}_j\right),
\end{eqnarray}
\begin{eqnarray}
\label{eq:active-fock}
 \hat{G} = \left(\hat{J}^t_u - \frac{1}{2} \hat{K}^t_u\right) D^u_t,
\end{eqnarray}
\begin{eqnarray}
\label{eq:active-meanfield}
 \hat{\mit \Gamma}_t \vert \phi_t \rangle = 
 \hat{W}^u_w \vert \phi_v \rangle P^{vw}_{tu}, 
\end{eqnarray}
\begin{eqnarray}
\label{eq:fc-core}
 \hat{h}^{\rm FC}(t) = \hat{h}(t) + 
 \sum_j^{\rm f.c.} \left(2\hat{J}_j(0) - \hat{K}_j(0)\right),
\end{eqnarray}
where summation $j$ in Eqs.~(\ref{eq:core-fock}) and
(\ref{eq:fc-core}) are restricted within dynamical-core (${\rm d.c.}$)
and frozen-core (${\rm f.c.}$) subspaces, respectively.
The operators $\hat{f}$ and $\hat{G}$ are universal and Hermitian, while
$\hat{\mit \Gamma}_t$ is defined with an active orbital $\phi_t$ to
be applied from the left, and non-Hermitian. We define Coulomb $\hat{J}$,
exchange $\hat{K}$, and general $\hat{W}$ mean field operators as
$ \hat{J}_p = \hat{J}^p_p, \hat{K}_p = \hat{K}^p_p, \hat{J}^p_q =
\hat{W}^p_q,  \hat{K}^p_q\vert\phi_r\rangle = \hat{W}^p_r \vert\phi_q\rangle$,
where $\hat{W}^p_q$ is local \cite{Kato:2008} and given in the coordinate space as 
\begin{eqnarray}
\label{eq:g-meanfield}
 W^p_q(\bm{r}) =
  \int\!d\bar{\bm{r}} \phi^*_p(\bar{\bm{r}}) 
  V_{ee}(\bm{r},\bar{\bm{r}}) \phi_q(\bar{\bm{r}}).
\end{eqnarray}
In Eq.~(\ref{eq:fc-core}), time argument $t$ is explicitly attached to
emphasize that the  time-dependence of the frozen-core dressed one-electron
Hamiltonian $\hat{h}^{\rm FC}(t)$ comes entirely from the external laser field
contribution in $\hat{h}(t)$.
%
Now Eq.~(\ref{eq:qg-tdmcscf}) for the time derivative matrix
$R^\mu_\nu={\rm i}\langle \phi_\mu \vert \dot{\phi}_\nu \rangle$ can be
worked out for inter-subspace (non-redundant) elements: 
\begin{eqnarray}\label{eq:r_ifc}
\begin{array}{ll}
R^\mu_i = R^{i*}_\mu = 0 & (i \in {\rm f.c.}),
\end{array}
\end{eqnarray}
\begin{eqnarray}\label{eq:r^a_idc}
\begin{array}{ll}
R^a_i = R^{i*}_a =
f^a_i + G^a_i & (i \in {\rm d.c.}),
\end{array}
\end{eqnarray}
\begin{eqnarray}\label{eq:r^a_t}
 R^a_t = R^{t*}_a =
  f^a_t +
  \left( {\mit\Gamma}_u \right)^a_u \left(D^{-1}\right)^u_t,
\end{eqnarray}
\begin{eqnarray}\label{eq:r^t_idc}
\begin{array}{c}
 R^t_i = R^{i*}_t =
 f^t_i + \left(\bar{D}^{-1}\right)^t_u
 \left\{ G^u_i - \left({\mit \Gamma}_u\right)^u_i \right\} \\
 (i \in {\rm d.c.}),
\end{array}
\end{eqnarray}
with $\bar{D}^t_u = 2\delta^t_u - D^t_u$, and for intra-subspace
(redundant) elements:
\begin{eqnarray}
\label{eq:r_redundant}
R^\mu_\nu = \langle \phi_\mu \vert \hat{\theta}(t) \vert \phi_\nu
\rangle,
\end{eqnarray}
where $\hat{\theta}(t)$ can be an arbitrary one-electron Hermitian operator
\cite{Caillat2005PRA, Kato:2004, Miranda:2011a}, reflecting the
invariance of the total wavefunction against the redundant orbital
transformations.

One could, in principle, directly work with
Eqs.~(\ref{eq:r_ifc})--(\ref{eq:r_redundant}) in the matrix formulation
\cite{Kato:2004}, which determines the time dependence of occupied
$\left\{\phi_p(t)\right\}$, as well as virtual
$\left\{\phi_a(t)\right\}$ orbitals. 
However, it is beneficial to introduce the orbital projector 
$\hat{Q} = \sum_a \vert \phi_a \rangle\langle \phi_a \vert
= 1 - \sum_p \vert \phi_p 
\rangle\langle \phi_p \vert$ onto the virtual orbital space, to avoid
(using the assumed completeness) explicitly dealing with numerous 
virtual orbitals 
\cite{Caillat2005PRA, Beck:2000}. 
Thus we arrive at the final
expression of EOMs for dynamical-core and active orbitals as follows:
\begin{eqnarray}
\label{eq:td-casscf-core}
 {\rm i} \vert\dot{\phi}_{i}\rangle =
  \hat{Q} \left( \hat{f} + \hat{G} \right) \vert\phi_i\rangle +
  \vert\phi_p\rangle R^p_i,
\end{eqnarray}
\begin{eqnarray}
\label{eq:td-casscf-active}
 {\rm i} \vert\dot{\phi}_t\rangle =
  \hat{Q} 
  \left\{ \hat{f} \vert\phi_t\rangle + 
   \hat{\mit\Gamma}_u \vert\phi_u\rangle \left(D^{-1}\right)^u_t \right\}
  + \vert\phi_p\rangle R^p_t,
\end{eqnarray}
and $R$ is determined by Eqs.~(\ref{eq:r^t_idc}) and
(\ref{eq:r_redundant}) with a particular choice of
$\hat{\theta}(t)$. Solving these equations guarantees the optimal
separation, in the TDVP sense, of frozen-core, dynamical-core,
active, and virtual orbital subspaces, as illustrated in
Fig.~\ref{fig:td-casscf}. 
This ensures the gauge-invariance of the TD-CASSCF method,
since the orbital subspaces are stable against single excitations
[Eq.~(\ref{eq:qg-tdmo})] arising with the transformation, e.g., from the
length gauge to the velocity gauge.

\subsubsection{CI equations of motion
\label{subsec:td-casscf-ci}}
The general CI-EOM of Eq.~(\ref{eq:g-tdci}) is specialized to the
TD-CASSCF method as
\begin{eqnarray}
\label{eq:td-casscf-ci}
{\rm i}\dot{C}_I = \sum_J^{\Pi_{\rm CAS}} 
\left(H^{\rm A}_{IJ} - \delta_{IJ}E^{\rm A} - R_{IJ}
\right) C_J, 
\end{eqnarray}
where $R_{IJ}=\langle I\vert\hat{R}\vert J\rangle$,
$E^{\rm A}$ and $H^{\rm A}_{IJ}$ are active orbital contributions to
the total energy and determinant basis Hamiltonian matrix elements,
respectively,
\begin{eqnarray}
\label{eq:cas-energy} 
E \equiv \langle \Psi \vert \hat{H} \vert \Psi \rangle = 
E^{\rm C} + E^{\rm A},
\end{eqnarray}
\begin{eqnarray}
\label{eq:hamiltonian-casci} 
\langle I\vert\hat{H}\vert J\rangle = \delta_{IJ} E^{\rm
C} + H^{\rm A}_{IJ}, 
\end{eqnarray}
where
\begin{eqnarray}
\label{eq:core-energy} 
E^{\rm C} = \sum_i^{\rm d.c.} \left\{\left(h^{\rm FC}\right)^i_i + 
f^i_i \right\},
\end{eqnarray}
\begin{eqnarray}
\label{eq:active-energy} 
E^{\rm A} = f^t_u D^u_t + \frac{1}{2} g^{tv}_{uw} P^{uw}_{tv},
\end{eqnarray}
\begin{eqnarray}
\label{eq:hamiltonian-casci-active} 
H^{\rm A}_{IJ} =
f^t_u (D_{IJ})^u_t + \frac{1}{2}
g^{tv}_{uw} (P_{IJ})^{uw}_{tv},
\end{eqnarray}
where $(D_{IJ})^u_t = \langle I\vert \hat{E}^t_u \vert J\rangle$, and
$(P_{IJ})^{uw}_{tv} = \langle I\vert \hat{E}^{tv}_{uw} \vert J\rangle$.
In Eq.~(\ref{eq:td-casscf-ci}), we make a particular phase choice,
${\rm i} \langle \Psi \vert \dot{\Psi} \rangle = 0$, by extracting the dynamical
phase $\exp\!\left[-{\rm i} \int^t dt^\prime E(t^\prime)\right]$ from the
total wavefunction. This stabilizes the CI-EOM especially when we have a
large active space.

\subsection{Computational remarks
\label{subsec:theory_remarks}}
The TD-CASSCF method includes as special cases both
TDHF and MCTDHF methods, thus bridges the gap between the uncorrelated and
fully correlated descriptions in a flexible way.
A practical advantage of this generality is that
a computational code written for the TD-CASSCF method can be used also for 
single-determinant TDHF and MCTDHF calculations, by
setting \{$n_{\rm C}=n$, $n_{\rm A}=0$\} and $\{n_{\rm C}=0$, $n_{\rm
A}=n$\}, respectively.
It can also execute open-shell TDHF calculation with fixed
CI coefficients \cite{Miranda:2011a}.
One indeed finds close similarity between TD-CASSCF EOMs
Eqs.~(\ref{eq:td-casscf-core}--\ref{eq:td-casscf-ci}) and those of the
MCTDHF method (See e.g. Ref.~\cite{Hochstuhl:2011}). Naively, ingredients
of the TD-CASSCF EOMs are the compilation of those for TDHF (core
orbitals) and MCTDHF (active orbitals and CI coefficients) methods.
This means that an existing code for the MCTDHF method can be easily generalized
to the TD-CASSCF method.



The computationally most demanding procedures required to integrate the
TD-CASSCF EOMs are grouped into two categories: 
\begin{enumerate}
 \item{Calculations of 2RDM elements $P^{tu}_{vw}$, and the two-electron
      contributions of Eq.~(\ref{eq:td-casscf-ci}),
      \begin{eqnarray}
       \label{eq:td-casscf-ci-2e} 
	{\rm i} \dot{C}_I \longleftarrow
	\frac{1}{2}\sum_J^{\Pi_{\rm CAS}}
	g^{tv}_{uw} (P_{IJ})^{uw}_{tv} C_J.
      \end{eqnarray}
      The amount of work in these procedures roughly scales as
      $O(N^2_{\rm A} (n_{\rm A} - N_{\rm A})^2 N_{\rm det})$ if
      $N^\alpha_{\rm A}=N^\beta_{\rm A}$, (see
      Ref.~\cite{Olsen:1988} for more details), where 
      $N_{\rm det}$ is the number of determinants in $\Pi_{\rm CAS}$
      which in turn scales factorially with the number of active
      electrons $N_{\rm A}$.}
 \item{Calculations of the mean fields $W^p_q(\bm{r})$, two-electron
      integrals $g^{tv}_{uw}$, and the 2RDM contributions in
      Eq.~(\ref{eq:td-casscf-active}),
      \begin{eqnarray}
       \label{eq:td-casscf-active-2rdm} 
	{\rm i} \dot{\phi}_t \longleftarrow
	\hat{W}^v_w \vert\phi_x\rangle P^{xw}_{uv} \left(D^{-1}\right)^u_t.
      \end{eqnarray}
      The computational cost of these steps depend explicitly on the
      number of grid points (or basis functions) $N_b$, as
      $O(n^2 N^2_b)$ for the mean fields and $O(n^4_{\rm A} N_b)$ for the
      others.
      }
\end{enumerate}

Important cost reductions are achieved for both procedures (A) and (B) by 
the TD-CASSCF method adopting core orbitals, compared to the MCTDHF method
with the same number of occupied orbitals. 
The speed-up and resource savings for procedure (A) is substantial due to
the decreased CI dimension.
This is especially the case if $N_{\rm A} \ll N$, which is expected
for an electronic structure with a few weakly-bound valence and
large numbers of physically inactive core electrons.
The cost reduction for procedure (B) is not as drastic as for (A), since
the amount of arithmetics $O(n^2 N^2_b)$ of
computing mean fields
$W^p_q(\bm{r})$ is independent of the CAS structure (only related to $n$). 
The computations of two electron integrals and
Eq.~(\ref{eq:td-casscf-active-2rdm}) become much faster through
restricting the orbital indices within the active instead of all occupied orbitals.  
Relative importance of these bottlenecks largely depends on the
problem at hand, and on the spatial representation of the orbitals and
electron-electron interactions. This point will be discussed in
Sec.~\ref{subsec:1d-lihd}.

\section{Numerical results and discussions\label{sec:results}}
In this section, we apply the TD-CASSCF method to the ionization
dynamics of one-dimensional (1D) multielectron model molecules.
The effective 1D Hamiltonian for $N$ electrons in the potential of $M$
fixed nuclei interacting with an external laser electric field $E(t)$ is taken as
\begin{widetext}
\begin{eqnarray}
\label{eq:1d-ham}
H = \sum_i^N \left\{-\frac{1}{2}\frac{\partial^2}{\partial x^2_i}
- \sum_a^M \frac{Z_a}{\sqrt{(x_i - X_a)^2 + c}}
- E(t)x_i
\right\} + \sum_{i>j}^N \frac{1}{\sqrt{(x_i-x_j)^2 + d}},
\end{eqnarray} 
\end{widetext}
where $x_i\, (i=1,\cdots,N)$ is the position of the $i$-th electron, $\bm{X}=\{X_a\}$ and
$\bm{Z}=\{Z_a\}$ $(a=1,\cdots,M)$ are the 
positions and charges of nuclei, and $c$ and $d$ adjust the soft Coulomb
operators of electron-nuclear and electron-electron interactions, respectively. The
electron-laser interaction is included within the dipole approximation
and in the length gauge.
Note that the TD-CASSCF method is gauge-invariant as
mentioned in Sec.~\ref{subsec:td-casscf-mo}. We have performed some of the
calculations described below also in the velocity gauge, and confirmed that
the results are virtually identical to those in the length gauge.

In this work, we make the simplest choice of $\hat{\theta}(t) \equiv 0$ in
Eqs.~(\ref{eq:r_redundant}), and therefore $R_{IJ} = 0$ in
Eq.~(\ref{eq:td-casscf-ci}). 
The orbital-EOMs are discretized on an equidistant
grid of spacing $\Delta x = 0.4$ (finer grid with $\Delta x
= 0.1$) is used for drawing Figs.~\ref{fig:lihd-pec}-\ref{fig:lihd-rho-orb}),
within a simulation box $\vert x| < L$ with $L = 600$. 
An absorbing boundary is implemented by the
mask function of $\cos^{1/4}$ shape at 15\% side edges of the box. The
ground-state electronic structure is obtained by the imaginary time
propagation with the fourth-order Runge-Kutta (RK4) algorithm with
Schmidt orthonormalization of orbitals after each
propagation \cite{Kato:2004}. The real-time propagations use variable
step-size embedded fourth- and fifth-order Runge-Kutta (VRK5) method. 
The kinetic energy operator
$-\frac{1}{2}\frac{\partial^2}{\partial x^2_i}$ is evaluated by the 
eighth-order finite difference, and spatial integrations are replaced by
grid summations using the trapezoidal rule. Further details of the
computations are given separately below.

\subsection{1D-LiH and LiH dimer models: Ground-state\label{subsec:ground-state}}

We consider 1D lithium hydride (LiH) and LiH dimer models.
The reason for choosing these models is that they
represent the simplest examples of such electronic structures with 
 (i) deeply bound orbitals, and
 (ii) {\it several} weakly bound orbitals, 
as shown below.
These characteristics should be the key in the three-dimensional
(3D) multielectron dynamics, where the existence of energetically
closely-lying valence electrons is quite common, which requires to
take both multichannel and multielectron effects into account.
As discussed previously \cite{Jordan:2006}, cares have to be made for the physical
soundness of 1D models. Nevertheless, we expect that the features (i) and (ii) are
transferable, and 1D applications can elucidate advantages and
limitations of theoretical methods, before applied to real 3D systems.

\begin{figure}[!t]
\includegraphics[width=0.5\textwidth,clip]{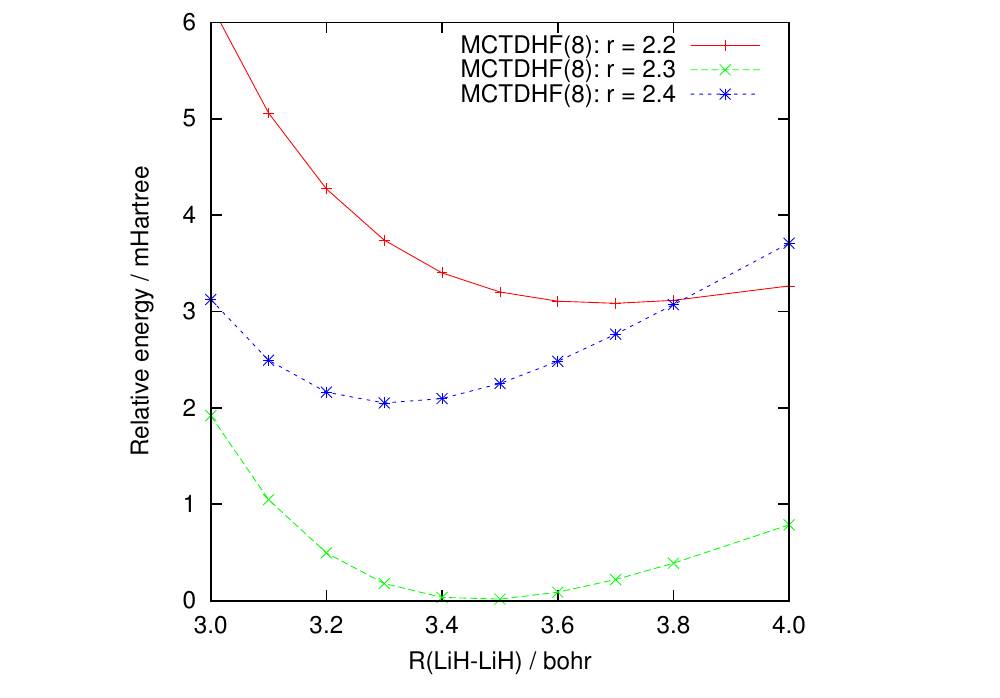}
\caption{\label{fig:lihd-pec}
 Several cuts of the adiabatic energy surface $E(r,R)$ of
 the 1D-(LiH)$_2$ model obtained by the MCTDHF($8$) method. The total energy of
 Eq.~(\ref{eq:bo-energy}) is plotted  
 against intermolecular LiH$\cdot\cdot\cdot$LiH distance $R$ for several
 bond lengths $r$ of LiH, constrained to be the same for two LiH molecules.}
\end{figure}

For LiH, we
set molecular parameters as $\bm{Z} = \{3, 1\}$ and $\bm{X} = \{-1.15, 1.15\}$.
For (LiH)$_2$, $\bm{Z}=\{3,1,3,1\}$ and $\bm{X} = \{-4.05, -1.75, +1.75,
+4.05\}$.
The soft Coulomb parameters $c = 0.5$ and $d = 1$ are used,
since an often-made choice of $c = d = 1$ \cite{Balzer:2010a, Balzer:2010b} was found to
overemphasize the electron-electron repulsion.
The above molecular parameters correspond to the equilibrium
bond length $r=2.3$ of Li-H and intermolecular distance $R=3.5$ of
LiH--LiH, as shown in Fig.~\ref{fig:lihd-pec}, which plots
several cuts of the adiabatic energy surface of (LiH)$_2$, 
\begin{eqnarray}
\label{eq:bo-energy} 
E(r, R)= \langle \Psi|H|\Psi\rangle + \sum_{a>b}^M \frac{Z_aZ_b}{|X_a-X_b|}.
\end{eqnarray}
The energy surface with parameters $c=d=1$ predicted no
stable LiH dimer in this nuclear configuration, relative to the
separated LiH molecules. 

To make a sensible comparison among methods with different
active spaces, we consider the following wavefunctions for LiH:
\begin{subequations}
\label{eq:cas_for_lih}
\begin{eqnarray}
\label{eq:cas_for_lih_hf}
\Psi_{{\rm HF}} &:& \phi^2_1 \phi^2_2, \\
\label{eq:cas_for_lih_cas2}
\Psi_{{\rm CASSCF}(2,n)} &:& \phi^2_1 \left(\phi_2 \phi_3
... \phi_{n+1}\right)^2, \\
\label{eq:cas_for_lih_mctdhf}
\Psi_{{\rm MCTDHF}(n+1)} &:& \left(\phi_1 \phi_2 \phi_3
... \phi_{n+1}\right)^4,
\end{eqnarray} 
\end{subequations}
and for (LiH)$_2$:
\begin{subequations}
\label{eq:cas_for_lihd}
\begin{eqnarray}
\label{eq:cas_for_lihd_hf}
\Psi_{{\rm HF}} &:& \phi^2_1 \phi^2_2 \phi^2_3 \phi^2_4, \\
\label{eq:cas_for_lihd_cas2}
\Psi_{{\rm CASSCF}(2,n-1)} &:& \phi^2_1 \phi^2_2 \phi^2_3
 \left(\phi_4 \phi_5 ... \phi_{n+2}\right)^2, \\
\label{eq:cas_for_lihd_cas4}
\Psi_{{\rm CASSCF}(4,n)} &:& \phi^2_1 \phi^2_2 
\left(\phi_3 \phi_4 ... \phi_{n+2}\right)^4, \\
\label{eq:cas_for_lihd_mctdhf}
\Psi_{{\rm MCTDHF}(n+2)} &:& \left(\phi_1 \phi_2 \phi_3
... \phi_{n+2}\right)^8,
\end{eqnarray}
\end{subequations}
following the notations of Eqs.~(\ref{eq:fullci_symbolic}) and (\ref{eq:casci_symbolic}).
The CASSCF and MCTDHF wavefunctions are designed to consist of the same
number of occupied orbitals with increasing active orbitals.

\begin{figure}[!t]
\includegraphics[width=0.4\textwidth,clip]{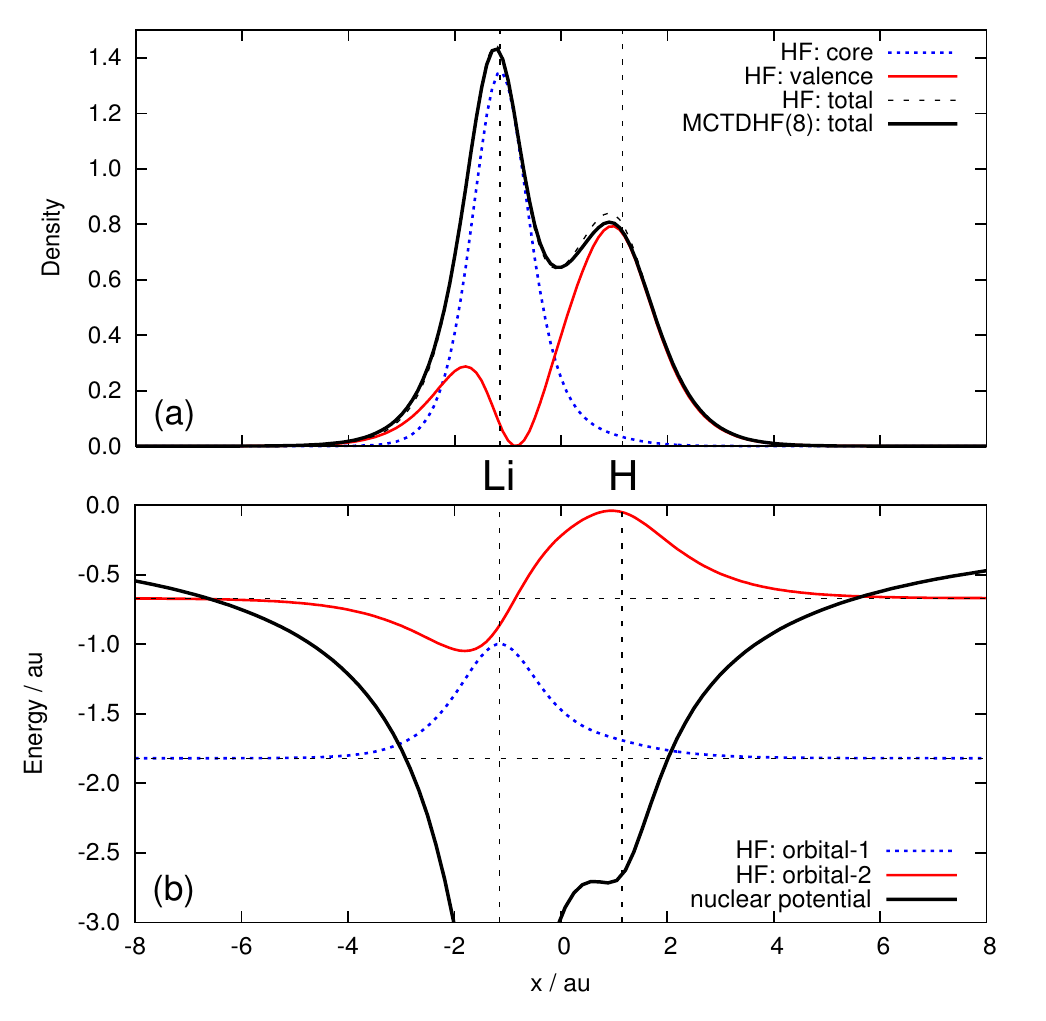}
\caption{\label{fig:lih-rho-orb}
 The ground-state electron density and occupied HF orbitals of the
 1D-LiH model. 
 (a) The total electron density by HF (black dashed) and MCTDHF($8$)
 (black thick solid) methods are
 compared. The two lines closely overlap each other. Also shown are the
 core (orbital 1) and valence (orbital 2) contributions to the total HF density.
 (b) The occupied HF orbitals are drawn in an arbitrary scale vertically
 shifted by orbital energies -1.82 (orbital 1) and -0.67 (orbitals
 2). The solid curve and dashed vertical lines show the nuclear
 potential and positions of nuclei, respectively.}%
\end{figure}

\begin{figure}[!t]
\includegraphics[width=0.4\textwidth,clip]{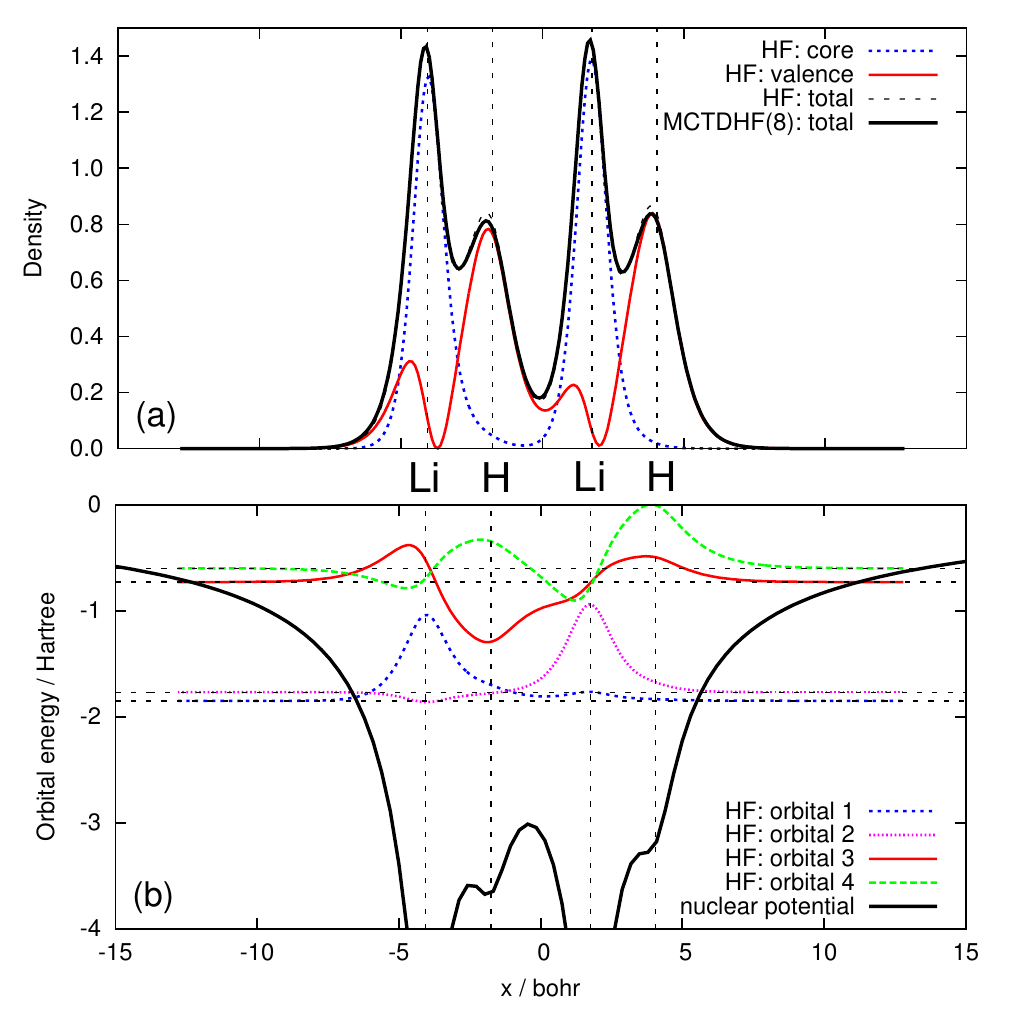}
\caption{\label{fig:lihd-rho-orb}
 Same as Fig.~\ref{fig:lih-rho-orb} for the 1D-(LiH)$_2$ model. 
 (a) The core and valence densities correspond to the contributions from
 orbitals 1 and 2, and orbitals 3 and 4, respectively.
 (b) The orbital energies are -1.85, -1.77, -0.73,  and -0.60, for
 orbitals 1 to 4, respectively.}%
\end{figure}

Figures ~\ref{fig:lih-rho-orb} and \ref{fig:lihd-rho-orb} show the shapes
of the ground-state occupied HF orbitals and the one-electron probability
density for LiH and (LiH)$_2$, respectively.
As seen in Fig.~\ref{fig:lih-rho-orb}(b) the
nodeless first deepest HF orbital of LiH localizes at Li ``atom'', while the
second orbital is responsible for the formation of ``chemical bond'',
made from constructive superposition of the ground-state wavefunction of H and the
second atomic orbital of Li, the node of the latter shifted to the
bonding region. Figure~\ref{fig:lih-rho-orb}(a) shows that the total
electron density is well reproduced by HF method compared to the
MCTDHF($8$) density. 
In Fig.~\ref{fig:lihd-rho-orb}(b),
one sees that HF orbitals of (LiH)$_2$ can be clearly separated to the
deeply-bound core (orbitals 1 and 2) and weakly-bound valence (orbitals
3 and 4) orbitals, the former keeping the atomic-orbital characters of
Li, while the latter two orbitals delocalizing across the dimer. Again,
as seen in Fig.~\ref{fig:lihd-rho-orb}(a), the total density is well
reproduced by HF.
Finally, it is observed that the tails of the total electron density
are determined by the valence electrons both in LiH and (LiH)$_2$.

Table~\ref{tab:lihd-ene} summarizes the ground-state calculations.
As seen in the table, there are significant gaps in total energies
between methods with different numbers of active electrons. However
the MCTDHF values for the other properties are reproduced rather well, by
CASSCF($2,n$) and CASSCF($4,n$) methods for LiH and (LiH)$_2$, respectively.
For instance, the difference in $E_{\rm tot}$ of CASSCF($4,8$) and
MCTDHF($10$) for (LiH)$_2$ is approximately 12 mHartree, but those in
IP and $\Delta_2$ are 0.05 eV (2 mHartree) and 0.003 eV (0.2
mHartree), respectively.
This tells that the correlations responsible for these
properties are those among the valence electrons.

The CASSCF($2,n$) active spaces of Eq.~(\ref{eq:cas_for_lihd_cas2})
, with only two of four nearly degenerate electrons being correlated,
are not physically sensible ones for (LiH)$_2$. Accordingly, the resulting dipole
moment values are not much improved from the HF value.
More seriously, the proper dissociation limit of such
wavefunction to the equivalent LiH molecules cannot be defined
well, i.e., the formation energy $\Delta_2$ cannot be obtained. This
problem is due to the lack of the ``size-extensivity'',
\cite{Szabo:1996, Schmidt:1998, Helgaker:2002}
which fails to guarantee the equal quality of the approximation for
different electronic configurations. The size-inextensive treatment
covers less and less electron correlation as systems grow larger.


\begin{table}[!b]
\caption{\label{tab:lihd-ene}
 Total energy $E_{\rm tot}$ in atomic unit (a.u.), dipole moment
 $\langle x\rangle$ in a.u., first ionization potential IP in eV, LiH
 formation energy $\Delta_1$ in eV, and (LiH)$_2$ formation energy
 $\Delta_2$ in eV. Results of HF, CASSCF, and  MCTDHF methods with
 varying active spaces are compared.
 The ionization potential (IP) is computed as the difference of the total
 energies of the neutral and cationic ground-states. 
 The $\Delta_1$ is the difference of the energy of
 LiH and the sum of energies of Li and H, and $\Delta_2$ is the
 difference of the energy of (LiH)$_2$ and twice that of
 LiH. For computing $\Delta_1$ and $\Delta_2$ with CASSCF and MCTDHF
 methods, the active spaces for the fragments are the proper
 dissociation limit of the parent description of the 
 complex. For HF calculations, the open-shell restricted HF
 method is used for doublet species.
 The $\Delta_2$ of CASSCF($2,n_{\rm A}$) methods are not available since
 their proper dissociation limits are not well defined.
}
\begin{ruledtabular}
\begin{tabular}{llddddd}
&&
\multicolumn{1}{c}{\textrm{$E_{\rm tot}$}} &
\multicolumn{1}{c}{\textrm{$\langle x\rangle$}} &
\multicolumn{1}{c}{\textrm{IP}} &
\multicolumn{1}{r}{$\Delta_1$} &
\multicolumn{1}{r}{$\Delta_2$} \\
\hline
\multicolumn{7}{c}{1D-LiH} \\
\multicolumn{7}{l}{$N_{\rm A}$ = 0} \\
 & HF          & -7.0664 & -0.133 & 17.82 & 5.07 & \\
\multicolumn{7}{l}{$N_{\rm A}$ = 2} \\
 & CASSCF(2,2) & -7.0819 & -0.141 & 18.24 & 5.49 & \\
 & CASSCF(2,4) & -7.0847 & -0.141 & 18.32 & 5.57 &\\
 & CASSCF(2,8) & -7.0847 & -0.141 & 18.32 & 5.57 &\\
\multicolumn{7}{l}{$N_{\rm A}$ = 4} \\
 & MCTDHF(3)   & -7.0824 & -0.141 & 18.15 & 5.51 &\\
 & MCTDHF(5)   & -7.0908 & -0.142 & 18.32 & 5.46 &\\
 & MCTDHF(9)   & -7.0920 & -0.142 & 18.35 & 5.48 &\\
\multicolumn{7}{c}{1D-(LiH)$_2$} \\
\multicolumn{7}{l}{$N_{\rm A}$ = 0} \\
 &  HF          & -14.1378 & -0.231 & 15.57 && 0.135  \\
\multicolumn{7}{l}{$N_{\rm A}$ = 2} \\
 &  CASSCF(2,3) & -14.1532 & -0.236 & 15.99 && \multicolumn{1}{r}{${\rm N/A}$} \\
 &  CASSCF(2,5) & -14.1534 & -0.236 & 16.00 && \multicolumn{1}{r}{${\rm N/A}$} \\
 &  CASSCF(2,7) & -14.1534 & -0.236 & 16.00 && \multicolumn{1}{r}{${\rm N/A}$} \\
\multicolumn{7}{l}{$N_{\rm A}$ = 4} \\
 &  CASSCF(4,4) & -14.1664 & -0.245 & 15.80 && 0.071 \\ %
 &  CASSCF(4,6) & -14.1726 & -0.246 & 15.90 && 0.100 \\ %
 &  CASSCF(4,8) & -14.1735 & -0.246 & 15.92 && 0.114 \\ %
\multicolumn{7}{l}{$N_{\rm A}$ = 8} \\
 &  MCTDHF(6)   & -14.1682 & -0.245 & 15.70 && 0.094 \\ %
 &  MCTDHF(8)   & -14.1822 & -0.248 & 15.85 && 0.110 \\ %
 &  MCTDHF(10)  & -14.1859 & -0.249 & 15.87 && 0.117 \\ %
\end{tabular}
\end{ruledtabular}
\end{table}

\subsection{1D LiH model: Ionization dynamics\label{subsec:1d-lih}}

Now we apply the TD-CASSCF method to the laser-driven electron
dynamics of the 1D-LiH model. 
We use the three-cycle laser electric field of the following form;
\begin{eqnarray}
\label{eq:field} 
E(t) = E_0 \sin(\omega t) \sin^2\left(\pi \frac{t}{\tau}\right), 
\hspace{.5em} 0 \leq t \leq \tau
\end{eqnarray}
with $\omega = 0.06075$ (wavelength 750 nm), $\tau = 6\pi/\omega$, and
three different amplitudes $E_0$ = 0.0534, 0.107, and 0.151
corresponding to peak intensities $I_0$ = 1.0$\times$10$^{14}$, 
4.0$\times$10$^{14}$, and 8.0$\times$10$^{14}$ W/cm$^2$, respectively.
The Keldysh parameters are 1.30, 0.65, and 0.46, respectively, for the three
intensities.
In view of the ground-state electronic structure of
Fig.~\ref{fig:lih-rho-orb} and the above laser profile, one reasonably
expects that the dominant physical process involved is the tunneling
ionization from the highest occupied orbital in the static HF picture.
Hence, we can speculate that the two-active-electron description 
TD-CASSCF($2,n_{\rm A}$) is necessary and sufficient for the accurate
description of the dynamics, as will be confirmed below.

\begin{figure}[!b]
\includegraphics[width=0.4\textwidth,clip]{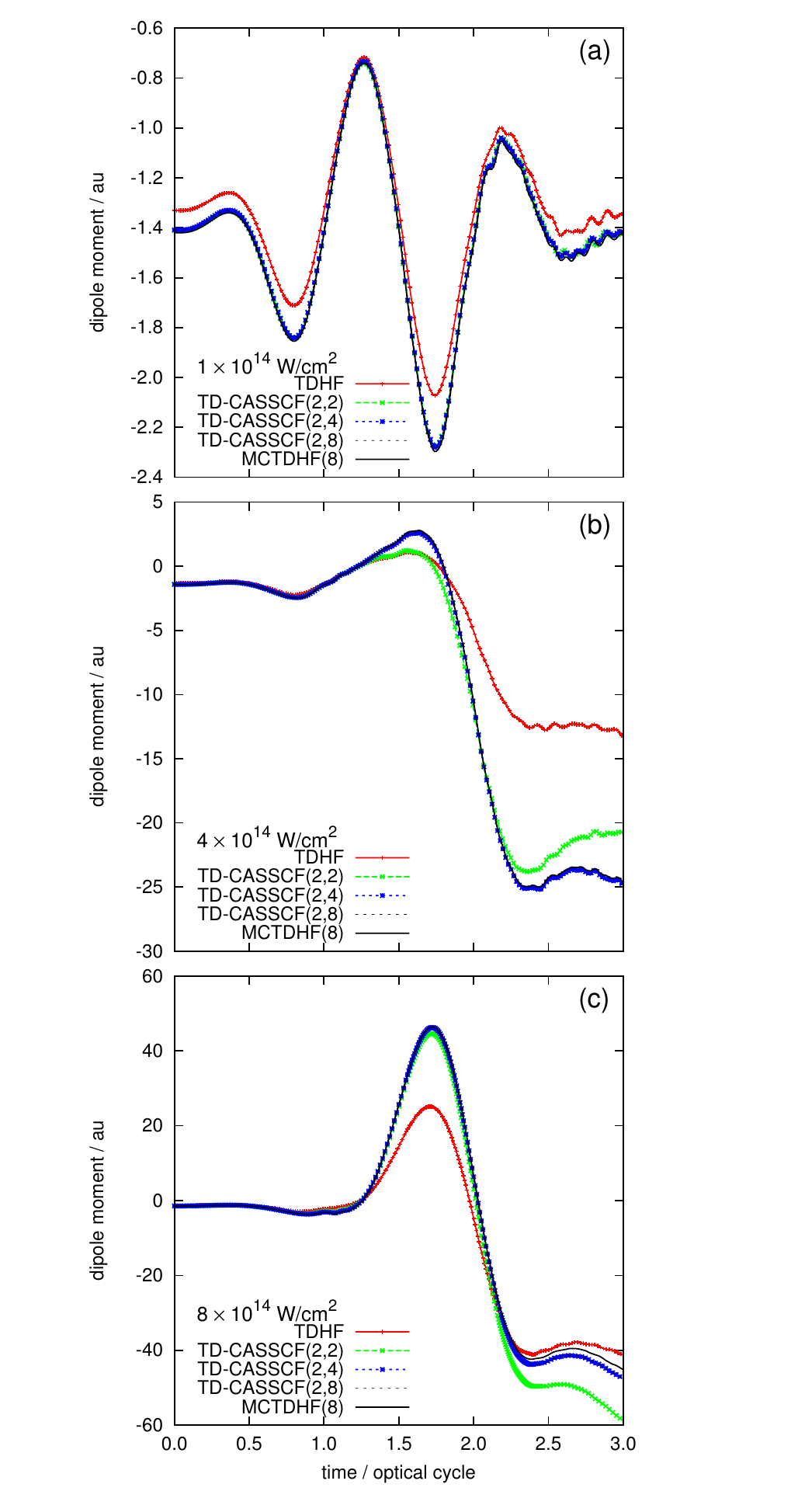}
\caption{\label{fig:lih-dip}
Dipole moment of the 1D-LiH model as a function of time, with peak intensities
(a) 1$\times$10$^{14}$, (b) 4$\times$10$^{14}$, and
(c) 8$\times$10$^{14}$ W/cm$^2$.}
\end{figure}

Figure~\ref{fig:lih-dip} shows the time evolution of the dipole
moment. 
First we observe the large difference in the results of TDHF and other
methods. For the lowest intensity of 
1.0$\times$10$^{14}$ W/cm$^2$, the difference remains quantitative,
largely due to the difference of the ground-state permanent dipole
moment. For higher intensities, TDHF clearly
underestimates the laser-driven large-amplitude electron motions.
This is due to the fundamental inadequacy of the closed-shell
description, Eq.~(\ref{eq:cas_for_lih_hf}), of the tunneling ionization
process, which involves spatially different motions of the ionizing and
non-ionizing electrons.
The TD-CASSCF($2,2$) brings substantial improvement over the TDHF, giving results
with much better agreement with the MCTDHF ones.
The convergent description in the TD-CASSCF($2,n_{\rm A}$) series is
obtained at $n_{\rm A} = 4$.
The TD-CASSCF($2,n_{\rm A}$) with $n_{\rm A} \geq 4$ closely
reproduce the results of MCTDHF method.

\begin{figure}[!b]
\includegraphics[width=0.5\textwidth,clip]{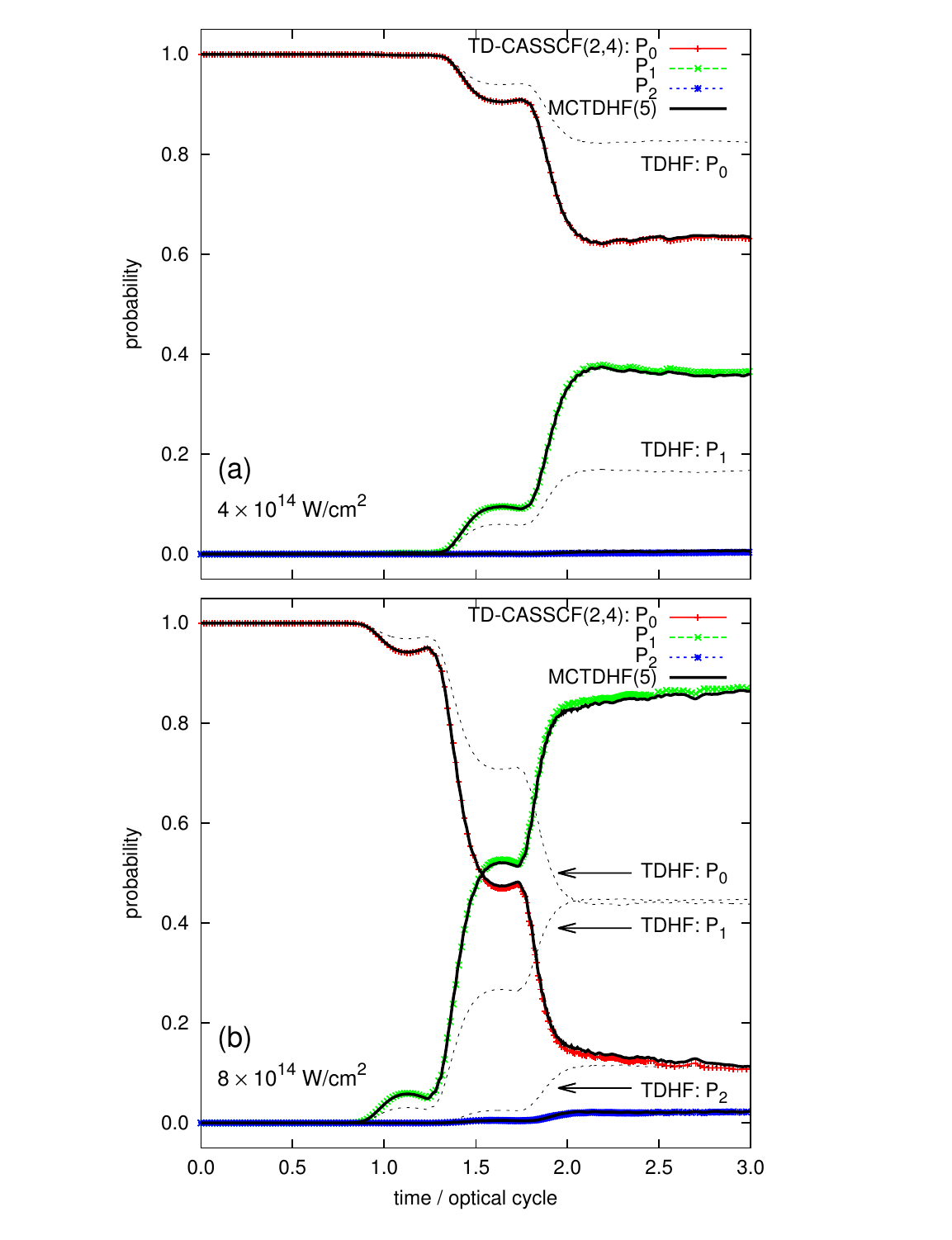}
\caption{\label{fig:lih-ipx}
Ionization probabilities $P_n$ of the 1D-LiH model as a function of
 time, with peak intensities 
 (a) 4$\times$10$^{14}$ and 
 (b) 8$\times$10$^{14}$ W/cm$^2$. TDHF (black dotted), TD-CASSCF($2,4$)
 (colored), and MCTDHF($5$) (block solid) results are compared.}
\end{figure}

Figure~\ref{fig:lih-ipx} plots the $n$-electron ionization
probability $P_n$, defined for convenience as 
a probability to find $n$ electrons located outside a given distance
$R_{\rm ion} = 20$
(see Appendix~\ref{app:ionp}),
of LiH as a function of time for the peak intensities (a) 4$\times$10\ue{14} and (b) 
8$\times$10\ue{14} W/cm\ue{2}. No appreciable ionization is found with
the lowest intensity. The probability of finding more than
two ionized electrons is negligibly small for all intensities. As seen in
Fig.~\ref{fig:lih-ipx}, TD-CASSCF($2,4$) gives virtually the same
results as MCTDHF($5$). The TDHF method, on the other hand, underestimates
single ionization $P_1$ and, at the higher intensity, unphysically
overestimates double ionization $P_2$. This is the consequence of
forcing two valence electrons to travel with a single spatial orbital.

These results demonstrate that the TD-CASSCF($2,2$) constitutes the simplest
method to describe the present dynamics in a physically correct way.
Its total wavefunction can be written as
\begin{eqnarray}
\label{eq:cas_lih_cas22}
\Psi
&=& \hat{A} \left[ \phi_1\bar{\phi}_1 \left\{ 
C_1 \phi_2\bar{\phi}_2 + C_2 \phi_3\bar{\phi}_3 \right\} \right] 
\end{eqnarray}
in the natural orbital representation
\cite{Kutzelnigg:1963, Kutzelnigg:1964}, 
where $\phi_i$ ($\bar{\phi_i}$) is
an orbital occupied by up (down) spin electrons. The two-configuration
CI part of Eq.~(\ref{eq:cas_lih_cas22}) can be transformed back to the
non-orthogonal expression [Eq.~(\ref{eq:gvb_ehf})],
\begin{eqnarray}
\label{eq:cas_lih_cas22_ehf}
\Psi
&\propto& \hat{A} \left[ \phi_1\bar{\phi}_1 \left\{ 
 (\psi_2\psi_3 + \psi_3\psi_2)
 \frac{1}{\sqrt{2}}(\alpha\beta - \beta\alpha) \right\} \right],
\end{eqnarray}
giving a clearer picture of different spatial motions of
the two valence electrons, with
\cite{Szabo:1996}
\begin{eqnarray}
\vert \langle \psi_1 \vert \psi_2 \rangle \vert = 
\frac{\vert C_1 \vert - \vert C_2 \vert}{\vert C_1 \vert + \vert C_2 \vert}. 
\end{eqnarray}
The flexibility inherent in Eqs.~(\ref{eq:cas_lih_cas22}) or
(\ref{eq:cas_lih_cas22_ehf}) enables a seamless transition from the
closed-shell dominant ground-state ($|C_1| \gg |C_2|
\Longleftrightarrow \langle\psi_1|\psi_2\rangle \approx 1$) to the
single ionization limit ($|C_1| \approx |C_2| \Longleftrightarrow
\langle\psi_1|\psi_2\rangle \approx 0$).
The ionization dynamics,
therefore, is characterized by the {\it strong} or {\it static} correlation
\cite{Roos:1987, Szabo:1996, Schmidt:1998, Helgaker:2002} 
in the sense that it involves drastic
changes of the configuration weights (the magnitudes of CI coefficients)
with more than one determinants contributing significantly.
The failure of single-determinant TDHF to describe the ionization
process is attributed to the lack of this type of correlation.

For quantitatively accurate description of the dynamics, 
the above minimum CI wavefunction has to be improved 
by incorporating more-than-two active orbitals, as seen in the convergence of the
dipole moments in Fig.~\ref{fig:lih-dip} with respect to the number of
active orbitals. The agreement of TD-CASSCF($2,n_{\rm A}$) and MCTDHF results
indicates that the core electron correlation is not relevant, at the first
approximation, for the ionization dynamics induced by the present laser
field. The TD-CASSCF allows the compact representation of such physical
situations.


\subsection{1D-LiH dimer model: Ionization dynamics\label{subsec:1d-lihd}}
In this section, we proceed to the multielectron dynamics of
1D-(LiH)$_2$ model.
We assess TDHF, TD-CASSCF($2,7$), TD-CASSCF($4,8$), and MCTDHF($10$)
methods. These active spaces are shown in Eq.~(\ref{eq:cas_for_lihd})
with $n = 8$. The latter two are twice the size of those in TD-CASSCF($2,4$)
and MCTDHF($5$) for LiH, respectively, which have been confirmed to
provide the convergent description in Sec.~\ref{subsec:1d-lih}.

\begin{figure}[!t]
\includegraphics[width=0.4\textwidth,clip]{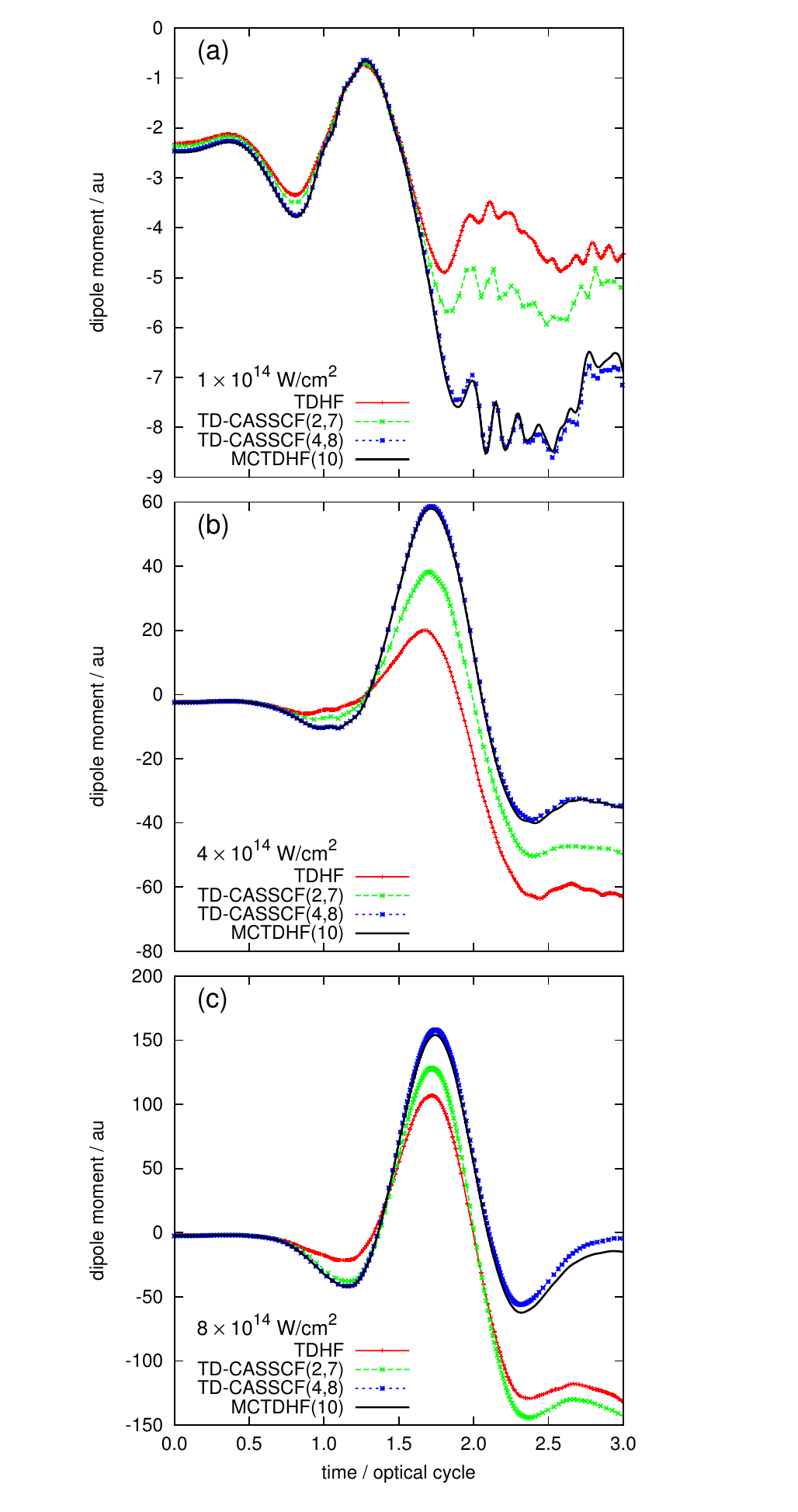}
\caption{\label{fig:lihd-dip}
Dipole moment of the 1D-(LiH)$_2$ model as a function of time, with peak intensities
(a) 1$\times$10$^{14}$, 
(b) 4$\times$10$^{14}$, and
(c) 8$\times$10$^{14}$ W/cm$^2$. Results of TDHF, TD-CASSCF($2,7$),
 TD-CASSCF($4,8$),  MCTDHF($10$) methods are compared.}
\end{figure}

Figure~\ref{fig:lihd-dip} shows the temporal evolution of the dipole
moment simulated with various methods. 
One clearly sees that TDHF and TD-CASSCF($2,7$) results show large
deviations from MCTDHF($10$) ones, while TD-CASSCF($4,8$) reproduces
the results of MCTDHF($10$) fairly well.
This indicates that all the four valence electrons sketched in
Fig.~\ref{fig:lihd-rho-orb} actively participate in the field-induced
ionization dynamics (this does not necessarily mean that the four
electrons are ionized),
while tightly bound core electrons remain non-ionized.
For the ionizing electrons, the closed-shell description is inadequate
as discussed in Sec.~\ref{subsec:1d-lih}.

\begin{widetext}

\begin{figure}[!t]
\includegraphics[width=0.9\textwidth]{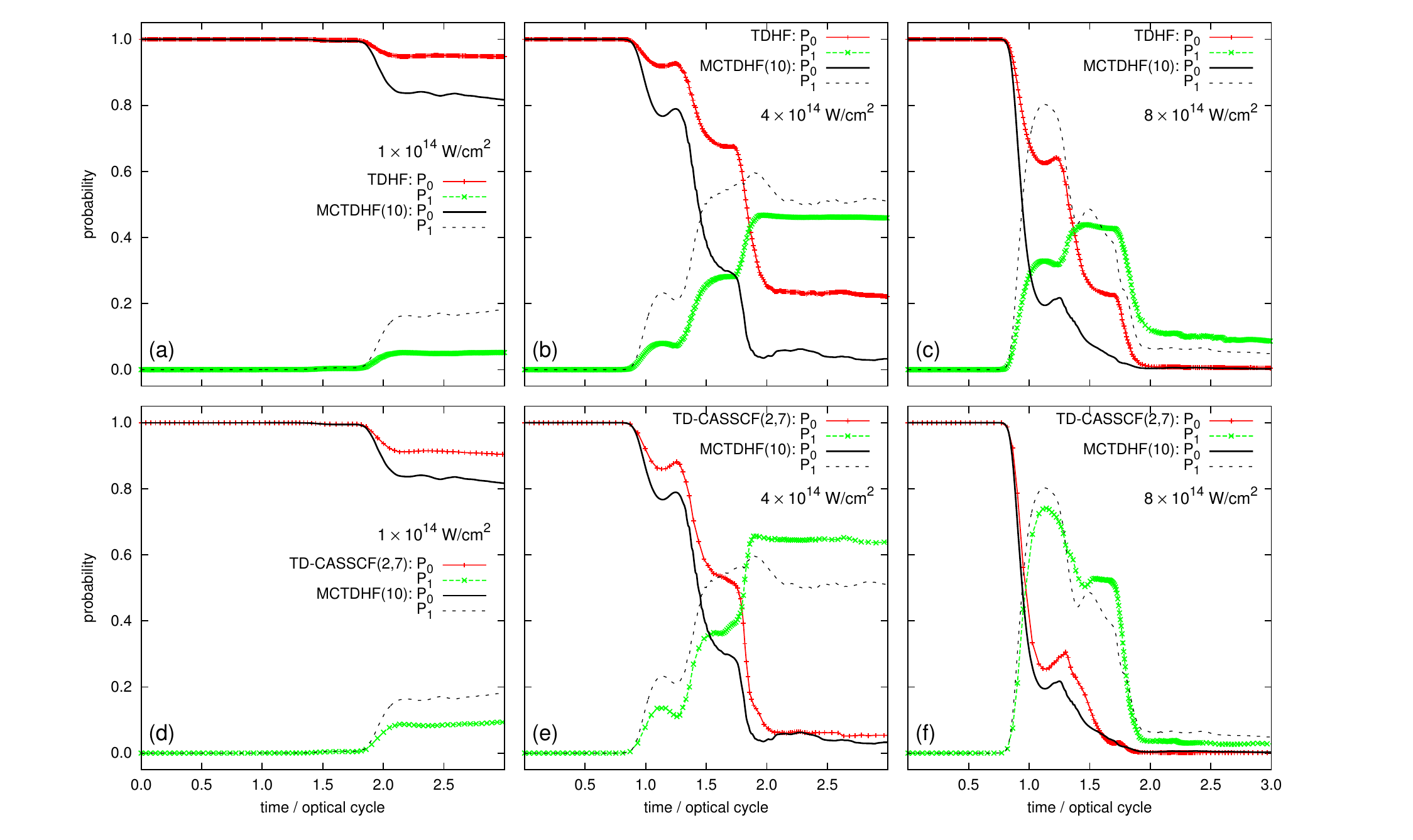}
\caption{\label{fig:lihd-ipx-1}
 Ionization probabilities $P_0$  and $P_1$ of the 1D-(LiH)$_2$ model as a function of
 time, with peak intensities 
 1$\times$10$^{14}$ (left),
 4$\times$10$^{14}$ (center), and 
 8$\times$10$^{14}$ W/cm$^2$ (right). Results of TDHF (top) and TD-CASSCF($2,7$) (bottom)
 are compared with those of MCTDHF (black solid and dotted).}
\end{figure}
\end{widetext}

\begin{widetext}

\begin{figure}[!t]
\includegraphics[width=0.9\textwidth,clip]{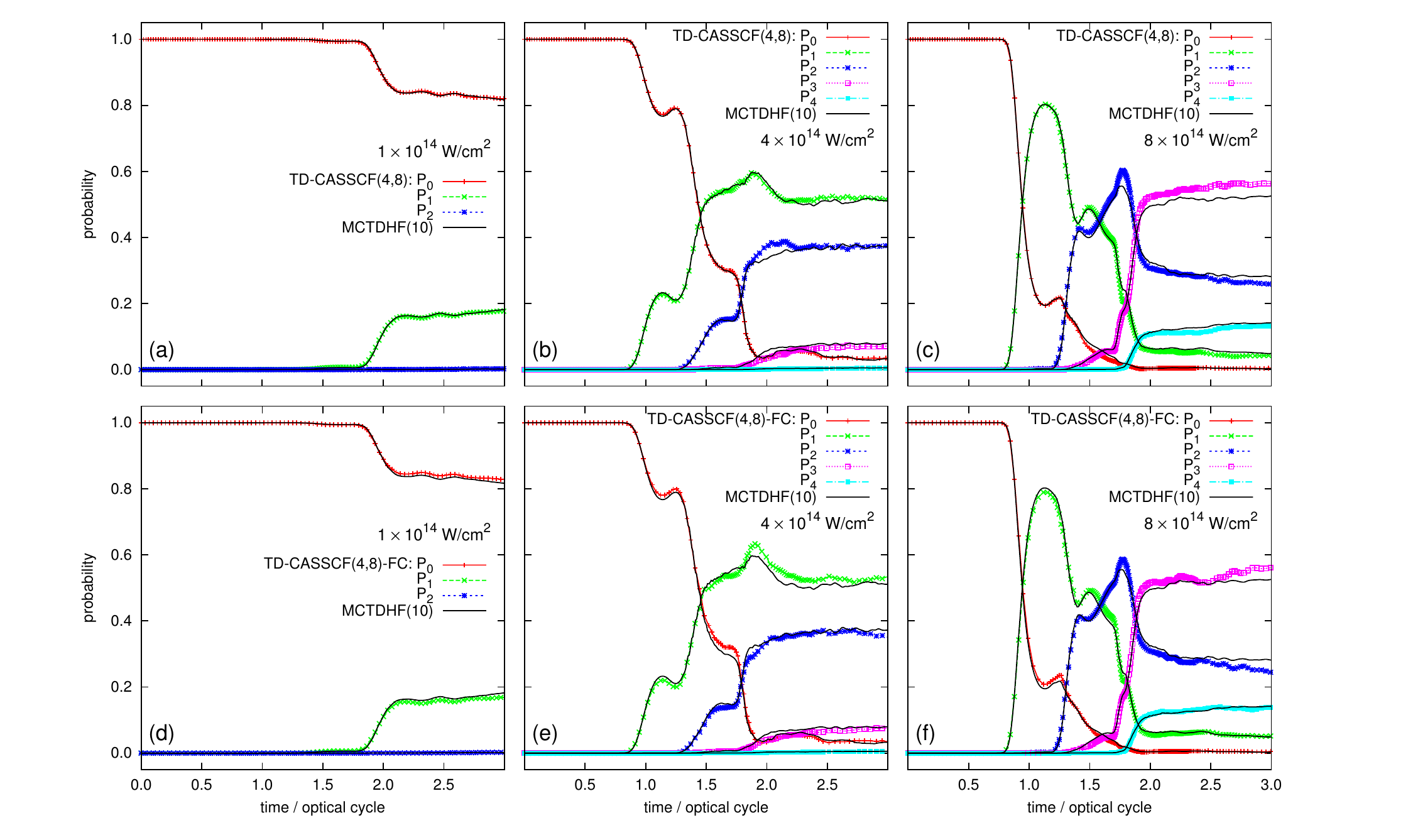}
\caption{\label{fig:lihd-ipx-2}
 Ionization probabilities $P_n$ with $n \leq 4$ of the 1D-(LiH)$_2$
 model as a function of
 time, with peak intensities 
 1$\times$10$^{14}$ (left),
 4$\times$10$^{14}$ (center), and 
 8$\times$10$^{14}$ W/cm$^2$ (right). Results of TD-CASSCF($4,8$) (top)
 and TD-CASSCF($4,8$)-FC (bottom) 
 are compared with those of MCTDHF (black solid).}
\end{figure}
\end{widetext}

In Figs.~\ref{fig:lihd-ipx-1} and \ref{fig:lihd-ipx-2}, we compare the
temporal evolution of the ionization probability $P_n$ with
$R_{\rm ion} = 20$ of (LiH)$_2$ computed by approximate 
methods and MCTDHF($10$). As can be seen from
Fig.~\ref{fig:lihd-ipx-1}, both TDHF and TD-CASSCF($2,7$) methods tend to
underestimate single ionization for all the examined intensities.
The probability of finding more than two ionized electrons are found to be
erroneous in an inconsistent way, thus not shown. In a striking contrast,
TD-CASSCF($4,8$) well reproduces the ionization probability 
$P_n$ obtained with MCTDHF($10$) [Fig~\ref{fig:lihd-ipx-2} (a)-(c)].
Slight deviation is seen only at the later stage of the pulse for the
higher intensities. The inclusion of more active orbitals would further
improve the agreement.

So far, all the core orbitals have been treated as dynamical core.
In Fig.~\ref{fig:lihd-ipx-2} (d)-(f), the
ionization probability computed with TD-CASSCF($4,8$) with all the core
orbitals treated as frozen, denoted TD-CASSCF($4,8$)-FC, are shown. It
reproduces the results of MCTDHF($10$) almost as nicely as
TD-CASSCF($4,8$) with dynamical-core orbitals, which indicates that the
core polarization plays minor roles in the present dynamics.

It is worth noting that even at the lowest intensity 1.0$\times$10$^{14}$
W/cm$^2$ dominated by single ionization, the TD-CASSCF($2,7$)
fails to give an accurate value of $P_1$ but underestimates it roughly by half [Fig.~\ref{fig:lihd-ipx-1} (d)]. This implies the importance of
the multichannel ionization, which can be described correctly only when
all the relevant orbitals are included in the active space. 
On the other hand, at higher intensities, 
the total wavefunction consists of the widespread superposition of the ground-,
excited- and continuum-states. For a balanced description, each of these
components has to be treated with an equal quality, which requires a
size-extensive theory. The MCTDHF, as the exact theory within a given
number of time-dependent bases, fulfills the size-extensivity condition. 
The TD-CASSCF with a proper active space preserves this important
property of the MCTDHF.
It is demonstrated by the accurate multiple ionization probabilities
obtained by the TD-CASSCF($4,8$) method, up to $P_4$ for the highest
intensity in Fig.~\ref{fig:lihd-ipx-2} (c).
The importance of selecting an appropriate active space is illustrated by the fact that
the TD-CASSCF($4,n_{\rm A}$) is required for (LiH)$_2$, while the
TD-CASSCF($2,n_{\rm A}$) is adequate for LiH.

\subsection{Analyses of computational cost\label{subsec:cost}}
Table~\ref{tab:lihd-cpu} summarizes computational times
for simulations of the 1D-(LiH)$_2$ model with a peak intensity
4$\times$10$^{14}$ W/cm$^2$.
To highlight the different computational bottlenecks discussed in
Sec.~\ref{subsec:theory_remarks}, several box sizes ($N_b$ = 1000, 2000,
and 3000) are considered.
%
The CPU times in table~\ref{tab:lihd-cpu}-(a), (b), and (c)
are recorded on a single Xeon processor of clock frequency 3.33 GHz 
for propagating 1000 time-steps during $2T \leq t \leq 2.1T$ 
with the fixed step-size RK4 algorithm, where $T = 2\pi/\omega$. 
Entry (d) compares wall clock times spent for 
completing the simulation up to four optical cycles ($0 \leq t \leq 4T$), with the VRK5
algorithm, measured for multi-threaded computations using 12 processors.

First, as seen in table~\ref{tab:lihd-cpu}-(a), CPU times for
procedure (A) grows rapidly with increasing CI dimension,
reproducing the theoretical linear dependence on $N_{\rm det}$.
These timings marginally depend on $N_b$.
Next, CPU times for procedure (B) in table~\ref{tab:lihd-cpu}-(b)
scale as $O(N^d_b)$ with $d$ = 1.95, 1.67, 1.66, and 1.47
for TDHF, TD-CASSCF($2,7$), TD-CASSCF($4,8$), and MCTDHF($10$) methods,
respectively. This is the consequence of competing $O(n^2 N^2_b)$
and $O(n^4_{\rm A}N_b)$ contributions as discussed in 
Sec.~\ref{subsec:theory_remarks}, with growing importance of
the latter for larger active spaces.
The TD-CASSCF($4,8$)-FC demands less CPU times than the TD-CASSCF($4,8$), due to
the strict locality of frozen-core orbitals, limiting the range of exchange
operators $\hat{K}_i|\phi_t\rangle$ around the core region.

Net CPU times are listed in table~\ref{tab:lihd-cpu}-(c). 
In TDHF and TD-CASSCF calculations with core subspaces, the
grid-intensive procedure (B) is definitely rate-limiting.
In contrast, MCTDHF calculations involve severe bottlenecks both in
procedures (A) and (B). The $N_{\rm det}$-dependent works
dominate 85\%, 67\%, and 55\% of the net CPU times, with $N_b$ = 1000,
2000, and 3000, respectively.
The cost reduction achieved by the TD-CASSCF method
largely depends on the relative importance of procedures (A) and (B).
Ratios of net CPU times for TD-CASSCF($4,8$) 
and MCTDHF($10$) calculations are 0.12, 0.29, and 0.45 with $N_b$ = 1000,
2000, 3000, respectively.
Similar trends are observed for wall clock times with VRK5
algorithm, as seen in table~\ref{tab:lihd-cpu}-(d). The stability of
EOMs is found to be similar for the tested methods, requiring about
70000 evaluations of EOMs.
The cost gain by the TD-CASSCF method relative to the MCTDHF method
will be more drastic if $N_{\rm A} \ll N$.
However, an efficient implementation of the mean field potential
[Eq.~(\ref{eq:g-meanfield})] is essential to achieve further speed-up
for large $N_b$, especially in three-dimensional applications.

\begin{table}[!h]
\caption{\label{tab:lihd-cpu}
 Computational times for simulations of the 1D-(LiH)$_2$ model.
 First row: Numbers of determinant $N_{\rm det}$ within the symmetry
 of zero spin projection.
 Entries (a), (b), and (c): Central processor unit (CPU) times in minutes
 for (A), (B), and overall procedures defined in
 Sec.~\ref{subsec:theory_remarks}, respectively.
 Entry (d): Wall clock times in minutes spent to complete the
 propagation of four optical cycles.
 Different simulation boxes are employed: $L = 200$
 ($N_b = 1000$), $L = 400$ ($N_b = 2000$), and $L = 600$
 ($N_b = 3000$). See text for more details.}
\begin{ruledtabular}
\begin{tabular}{lddddd}
&
\multicolumn{1}{c}{\textrm{TDHF}} &
\multicolumn{3}{c}{\textrm{TD-CASSCF}} &
\multicolumn{1}{c}{\textrm{MCTDHF}} \\
\cline{3-5}
active space&
\multicolumn{1}{c}{\textrm{(0,0)}} &
\multicolumn{1}{c}{\textrm{(2,7)}} &
\multicolumn{1}{c}{\textrm{(4,8)}} &
\multicolumn{1}{c}{\textrm{(4,8)-FC}} &
\multicolumn{1}{c}{\textrm{(8,10)}} \\
\hline
$N_{\rm det}$ & 1 & 49 & 784 & 784 & 44100 \\
\multicolumn{6}{l}{(a) RK4 / 1000 steps, CPU-A} \\
$N_b$ = $1000$ & 0.0 & 0.1 & 1.2 & 1.2 & 212.4 \\
$N_b$ = $2000$ & 0.0 & 0.2 & 1.3 & 1.3 & 215.1 \\
$N_b$ = $3000$ & 0.0 & 0.3 & 1.4 & 1.4 & 215.3 \\
\multicolumn{6}{l}{(b) RK4 / 1000 steps, CPU-B} \\
$N_b$ = $1000$ &  4.3 &  26.8 &  28.4 &  22.0 &  35.7 \\
$N_b$ = $2000$ & 16.3 &  87.6 &  90.9 &  73.8 & 103.5 \\
$N_b$ = $3000$ & 36.8 & 168.0 & 175.4 & 132.9 & 177.2 \\
\multicolumn{6}{l}{(c) RK4 / 1000 steps, CPU net} \\
$N_b$ = $1000$ &  4.4 &  27.0 &  29.8 &  23.4 & 248.9 \\
$N_b$ = $2000$ & 16.4 &  88.1 &  92.5 &  75.4 & 319.4 \\
$N_b$ = $3000$ & 36.9 & 168.7 & 177.3 & 134.7 & 393.5 \\
\multicolumn{6}{l}{(d) VRK5 / 4 cycles, Wall} \\
$N_b$ = $1000$ &  8.2 &  43.7 &  51.3 &  39.1 & 451.5 \\
$N_b$ = $2000$ & 31.0 & 174.8 & 192.6 & 141.4 & 628.8 \\
$N_b$ = $3000$ & 65.2 & 378.5 & 394.0 & 282.9 & 823.1 \\
\end{tabular}
\end{ruledtabular}
\end{table}

\section{Conclusions\label{sec:conclusions}}
We have developed a new {\it ab initio} time-dependent many-electron
method called TD-CASSCF. It applies the concept of CASSCF,
which has been developed for the electronic structure calculation in
quantum chemistry, to the multielectron dynamics in intense laser
fields, introducing frozen-core, dynamical-core, and active orbital
subspaces. The classification into the subspaces can be done
flexibly conforming to simulated physical situations and desired
accuracy, and both TDHF and MCTDHF methods are included as special
cases. This feature enables compact yet accurate 
representation of ionization dynamics in many-electron systems, and
bridge the huge gap between TDHF and MCTDHF methods.

We have applied the TD-CASSCF method to the
ionization dynamics of 1D-LiH and 1D-(LiH)$_2$, to assess its capability to
describe multichannel and multielectron ionization.
It has been confirmed that the present method closely reproduces rigorous MCTDHF
results if the active orbital space is properly chosen to include
appreciably ionizing electrons.
We have also confirmed that the TD-CASSCF provides substantial computational cost
reduction in the CI-length dependent procedures, which scale by far the
steepest with the system size in the MCTDHF method.
Therefore, the TD-CASSCF method is most advantageous
for problems in which only a few weakly-bound electrons 
out of a large number of total electrons ionize.

While it is sometimes stated that the MCTDHF method is 
a time-dependent version of the CASSCF method
\cite{Nest:2005a, Nest:2005b},
this statement is even more suitable for the TD-CASSCF method
introduced in the present study.
With reduced computational cost, the TD-CASSCF method with a properly
chosen active space preserves most of the theoretically important
properties of the MCTDHF:
(i) flexibility to account for the strong-correlation involved
in the ionization dynamics,
(ii) size-extensivity, essential for a balanced description of different
electronic configurations, 
(iii) gauge-invariance by virtue of the time-dependent variational
optimization of orbitals, and
(iv) invariance against orbital transformation within an orbital subspace, allowing
e.g., the natural orbital analyses of the time-dependent wavefunction \cite{Kato:2009}.


It should be noted that the computational cost of
the TD-CASSCF method still scales factorially with the number of active (not total)
electrons, thus its applications are limited to, say, 16 half-filled
active orbitals in view of the present state of the art in quantum
chemistry. An example requiring such a large active space is the
ionization from densely lying multiple valence orbitals in
weakly-interacting molecular clusters.
To approach to such a problem, more restricted (instead of complete)
constructions of the active space will be necessary.
Moreover, a breakthrough is needed to represent one-particle
wavefunctions in the general molecular potential without 
particular symmetries.
In spite of these challenges, we foresee that the TD-CASSCF method
will find fruitful applications in multielectron dynamics of, e.g.,
rare gas atoms heavier than helium, or 
molecules composed of atoms in the
first few rows of the periodic table, exposed to visible-to-mid-infrared 
high-intensity pulses, which are inaccessible with the
all-electron-active MCTDHF method.

 
\begin{acknowledgments}
 This research is supported in part by Grant-in-Aid for Scientific
 Research (No. 23750007, No. 23656043, and No. 23104708) from the
 Ministry of Education, Culture, Sports, Science and Technology (MEXT)
 of Japan, and also by Advanced Photon Science Alliance (APSA) project
 commissioned by MEXT.
\end{acknowledgments}

\appendix
\section{Calculation of ionization probabilities\label{app:ionp}}
To conveniently evaluate the multiple ionization yield in many electron systems, 
we introduce a domain-based ionization probability $P_n$, defined as a probability
to find $n$ electrons in the outer region $|\bm{r}| >
R_{\rm ion}$ and the remaining $N - n$ 
electrons in the inner region $|\bm{r}| < R_{\rm ion}$,
with a given distance $R$ from the origin,
\begin{eqnarray}
\label{eq:ionp} 
P_n &\equiv& \binom{N}{n} 
\int_> dx_1 \cdot\cdot
\int_> dx_n
\int_< dx_{n+1} \cdot\cdot
\int_< dx_N \nonumber \\ &&
\Psi^*(x_1,\cdot\cdot,x_N)* \Psi(x_1,\cdot\cdot,x_N),
\end{eqnarray} 
where $\int_<$ and $\int_>$ symbolize integrations over a spatial-spin
variable $x = \{\bm{r}, \xi\}$ with the spatial part restricted to the domains
$|\bm{r}| < R_{\rm ion}$, and $|\bm{r}| > R_{\rm ion}$, respectively.

It is convenient to introduce
an auxiliary quantity $T_n$ obtained by replacing the 
outer-region integrals in Eq.~(\ref{eq:ionp}) with the full-region ones ($\int_>
\rightarrow \int$). It relates to $P_n$ as 
\begin{eqnarray}
\label{eq:ionp-t}
P_n = \sum_{k=0}^n \binom{N-n+k}{k}(-1)^k T_{n-k}.
\end{eqnarray}
By adopting the CI expansion of Eq.~(\ref{eq:mcscf_2q}), and making use of
the orthonormality of spin-orbitals in the full-space integration, we
have 
\begin{eqnarray}
\label{eq:iont}
T_n &=& \sum_{IJ}^\Pi C^*_I C_J D^{(n)}_{IJ},
\end{eqnarray}
where
\begin{eqnarray} 
\label{eq:iond}
D^{(0)}_{IJ} &=& \sum_{ij}^N
\det\left(S^<_{IJ}\right), \nonumber \\
D^{(1)}_{IJ} &=& \sum_{ij}^N \epsilon^{IJ}_{ij} 
\det\left(S^<_{IJ}[i:j]\right), \nonumber \\
D^{(2)}_{IJ} &=& \sum_{i>j}^N \sum_{k>l}^N \epsilon^{IJ}_{ik}
 \epsilon^{IJ}_{jl}
\det\left(S^<_{IJ}[ij:kl]\right),
	       \end{eqnarray}
etc, and  $S^<_{IJ}$ is an $N\times N$
matrix with its $\{ij\}$ element being the inner-region overlap integral,
\begin{eqnarray}
\label{eq:overlap}
(S^<_{IJ})_{ij} = \int_< dx \phi^*_{p(i,I)}(x) \phi_{q(j,J)}(x) \equiv
\langle \phi_p | \phi_q \rangle_<,
\end{eqnarray}
with $\phi_{p(i,I)}$ being the $i$-th (in a predefined order) spin-orbital in
the determinant $I$. $S^<_{IJ}[ij\cdot\cdot:kl\cdot\cdot]$ is the
submatrix of $S^<_{IJ}$ obtained after removing rows $i,j,\cdot\cdot$ and columns
$k,l,\cdot\cdot$ from the latter, and
\begin{eqnarray}
\label{eq:epsilon}
\epsilon^{IJ}_{ij} = \delta^{p(i,I)}_{q(j,J)} (-1)^{i+j}.
\end{eqnarray}
The matrix $S^<_{IJ}$ and its submatrices are block-diagonal due to
the spin-orthonormality, so that, e.g., $\det\left(S^<_{IJ}\right) = 
\det\left(S^<_{I^\alpha J^\alpha}\right) 
\det\left(S^<_{I^\beta J^\beta}\right)$, where $I^\sigma$ is the
$\sigma$-spin part of the determinant $I$. 

The procedure given above remains a manageable task in the present
applications, up to eight (all) electron ionization probabilities in the
1D-(LiH)$_2$ model. While this scheme becomes impractical for
systems with more electrons, it may still be useful for problems where
only a few electrons are ejected appreciably, since the dimension of
Eqs.~(\ref{eq:iond}) can be reduced to the number of the ionizing
electrons.

This approach allows the evaluation of multiple ionization yields by
using the information of the inner region orbitals $\langle \phi_p |
\phi_q \rangle_<$ and the formal orthonormality relation $\delta^p_q =
\langle \phi_p | \phi_q \rangle_< + \langle \phi_p | \phi_q \rangle_>$.
It works with a reasonable size of the simulation box $L$,
provided that $R_{\rm ion} \ll L$ and a good absorber is implemented to
prevent the reflection of the wavefunction. 
In fact, we performed calculations for the 1D-(LiH)$_2$ model in
Sec.~\ref{sec:results} using smaller boxes with $L$ = 200 and 400 a.u., where a sizable
portion of the norm is lost at the boundary, and confirmed that the obtained
ionization yields are virtually the same with those of
Fig.~\ref{fig:lihd-ipx-1} and \ref{fig:lihd-ipx-2}. Such small-scale calculations
could serve as preliminary validations for the choice of the active
space before stepping into large-scale computations.

\end{document}